\documentclass[12pt]{article}

\usepackage[latin1]{inputenc}
\usepackage[T1]{fontenc}
\usepackage{epsfig} 
\usepackage{color} 
\usepackage{graphicx} 
\usepackage{geometry} 
\usepackage{placeins} 
\usepackage{textcomp}
\usepackage[numbers,sort&compress]{natbib}
\usepackage{times}

\newenvironment{sciabstract}{%
\begin{quote} \bf}
{\end{quote}}

\newcounter{lastnote}

\title{Methane storms as a driver of Titan's dune orientation}

\author
{Benjamin Charnay$^{1,2}$, Erika Barth$^{3}$, Scot Rafkin$^{3}$, Cl\'ement Narteau$^{4}$, \\ S\'ebastien Lebonnois$^{2}$, S\'ebastien Rodriguez$^{5}$, Sylvain Courrech du Pont$^{6}$ \\
and Antoine Lucas$^{5}$\\
\\
\\
\normalsize{$^{1}$Virtual Planetary Laboratory, University of Washington,}\\
\normalsize{Box 351580, Seattle, WA 98195, USA}\\
\normalsize{$^{2}$Laboratoire de M\'et\'eorologie Dynamique, IPSL/CNRS/UPMC, 75252 Paris, France}\\
\normalsize{$^{3}$Southwest Research Institute, Boulder, CO 80302, USA}\\
\normalsize{$^{4}$Institut de Physique du Globe de Paris, CNRS-UMR 7154,}\\
\normalsize{Sorbonne Paris Cit\'e, Universit\'e Paris-Diderot, 75238 Paris, France.}\\
\normalsize{$^{5}$Laboratoire Astrophysique, Instrumentation et Mod\'elisation,}\\
\normalsize{CNRS-UMR 7158, Universit\'e Paris-Diderot, CEA-SACLAY, 91191 Gif sur Yvette, France }\\
\normalsize{$^{6}$Laboratoire de Mati\`ere et Syst\`emes Complexes,}\\
\normalsize{CNRS-UMR 7057, Sorbonne Paris Cit\'e, Universit\'e Paris-Diderot, 75013 Paris, France}\\
\\
\normalsize{E-mail:  benjamin.charnay@lmd.jussieu.fr.}
}

\date{}

\begin{document} 

\baselineskip14pt

\maketitle

\begin{sciabstract}
Titan's equatorial regions are covered by eastward propagating linear dunes\citep{lorenz06, lorenz09, rodriguez14}. This direction is opposite to mean surface winds simulated by Global Climate Models (GCMs), which are oriented westward at these latitudes, similar to trade winds on Earth \citep{lorenz06, tokano08}. Different hypotheses have been proposed to address this apparent contradiction, involving Saturn's gravitational tides\citep{lorenz06}, large scale topography\citep{tokano08} or wind statistics \citep{tokano10}, but none of them can explain a global eastward dune propagation in the equatorial band.
Here we analyse the impact of equinoctial tropical methane storms developing in the superrotating atmosphere (i.e. the eastward winds at high altitude) on Titan's dune orientation. Using mesoscale simulations of convective methane clouds \citep{barth07,barth10} with a GCM wind profile featuring superrotation \citep{lebonnois12}, we show that Titan's storms should produce fast eastward gust fronts above the surface. Such gusts dominate the aeolian transport, allowing dunes to extend eastward. This analysis therefore suggests a coupling between superrotation, tropical methane storms and dune formation on Titan. Furthermore, together with GCM predictions and analogies to some terrestrial dune fields, this work provides a general framework explaining several major features of Titan's dunes: linear shape, eastward propagation and poleward divergence, and implies an equatorial origin of Titan's dune sand.

\end{sciabstract}

A major surprise in the exploration of Titan by Cassini was the discovery of large dune fields in the equatorial regions, which cover close to 15$\%$ of Titan's surface \citep{lorenz06, lorenz05, rodriguez14}. These giant dunes are linear and parallel to the equator, and are probably composed of  hydrocarbon material. The analysis of dune morphology around obstacles and dune terminations indicates an eastward dune propagation with some regional variations\citep{lorenz06, lorenz09} (see also Fig. \ref{figure_3}b and Supplementary Fig. 5). 
However, Global Climate Models (GCMs) predict that annual mean surface winds are easterlies (westward) at low latitudes \citep{tokano08}, as trade winds on Earth \citep{zhu08}. Therefore, Titan's dune orientation is opposite to predicted mean winds, raising a major enigma.

Given the non-linear dependence of sediment transport on wind speed, the only way to propagate dunes eastward would be the occurrence of episodic fast westerly (eastward) gusts \citep{tokano10}. Above 5 km, Titan's troposphere is in superrotation with fast eastward winds at any latitude. Pumping momentum from this superrotation down to the surface might provide a solution. However, Titan's tropospheric circulation is essentially confined into a 2 km boundary layer \citep{charnay12}. Incidentally, this boundary layer circulation may explain the dune spacing of around 2 km \citep{charnay12,lorenz10a}. Dry convection is therefore limited to the first 2 km and unable to reach the fast eastward winds \citep{charnay12}. The Koln GCM has been shown to generate episodic fast eastward gusts at the equator during Titan's equinoxes \citep{tokano10}. However, this result has not been reproduced by any published GCM, including the Titan IPSL GCM (used here), which faithfully reproduces the superrotation and the thermal structure measured by the Huygens probe in the troposphere \citep{lebonnois12, charnay12} in contrast to the Koln GCM.
We therefore conclude that Titan's atmospheric general circulation is not likely to produce fast eastward winds at the equator.

Similar to the water cycle on Earth, Titan's weather is characterized  by a methane cycle producing methane clouds. These clouds are rare over the equatorial region and probably absent during most of Titan's year \citep{rodriguez11}. 
However, during the equinox, the circulation leads to a methane convergence at the equator, sufficient to trigger deep convection \citep{mitchell11,schneider12}. Tropical convective clouds have been detected during the equinoctial season \citep{turtle11b, turtle11a, rodriguez11,griffith09} with top altitude between 10 and 30 km \citep{griffith09}.
Huge storm systems covering a large part of the equatorial band have been observed, one of them associated with methane precipitations \citep{turtle11b,turtle11a}. 
On Earth, storms frequently generate gust fronts above the surface. They occur when downdrafts, generally cooled by rain evaporation, reach the ground below the storm, producing cold surface currents, also called cold pools \citep{mahoney09}. The gust front is the leading edge of the cold pool, and its propagation is mostly controlled by the direction of upper winds.

To investigate whether Titan's storms might lead to the production of such gusts with a preferential direction, we analysed their formation under the wind-shear produced by Titan's superrotation with a mesoscale regional model.
We ran 2-dimensional (longitude-altitude) mesoscale simulations with TRAMS \citep{barth07,barth10}. This model includes full cloud microphysics (see Supplementary Information) with condensation, melting, freezing and evaporation of methane cloud particles.

\begin{figure}[h] 
\begin{center} 
	\includegraphics[width=7cm]{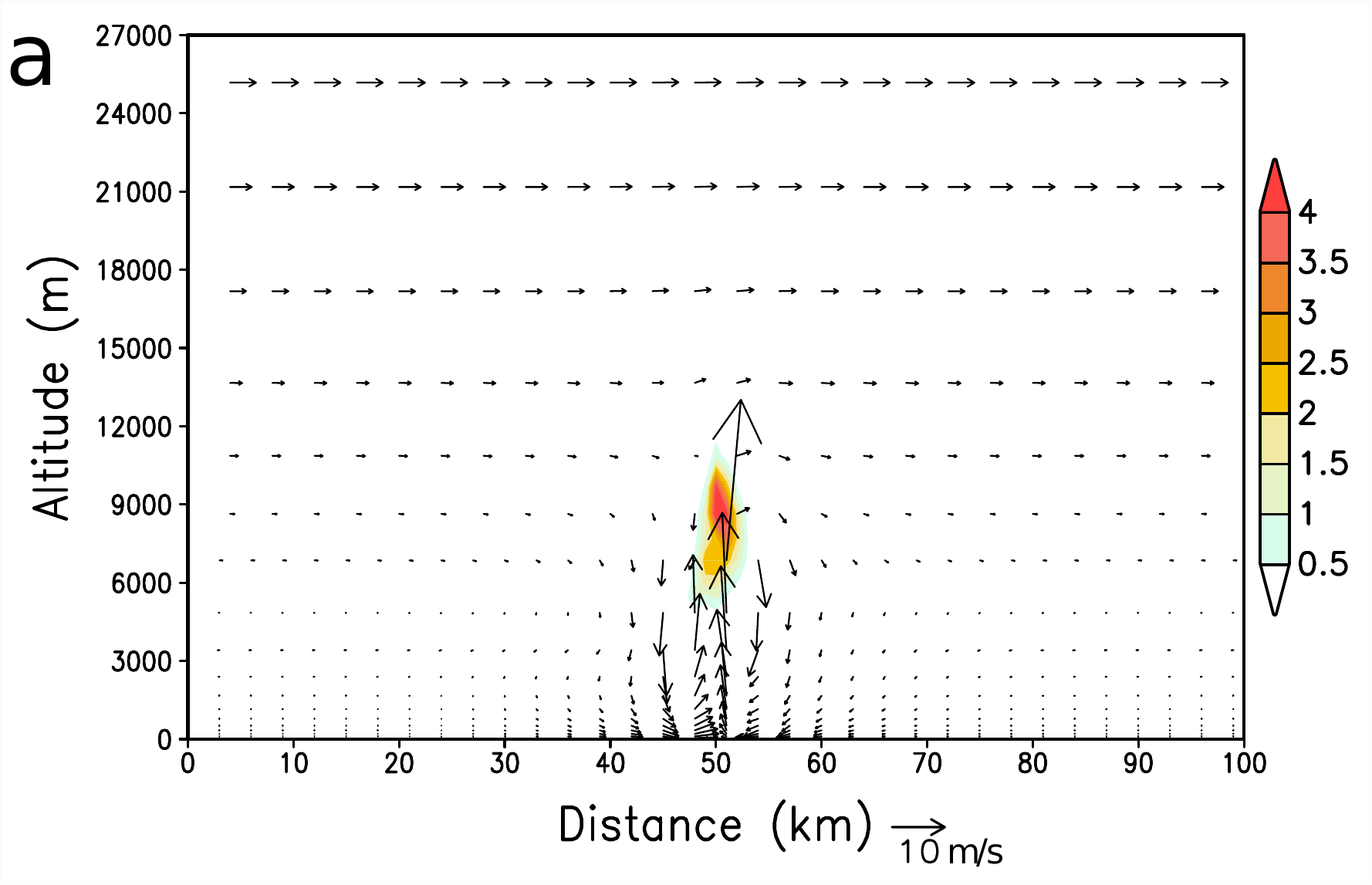}
        \includegraphics[width=7cm]{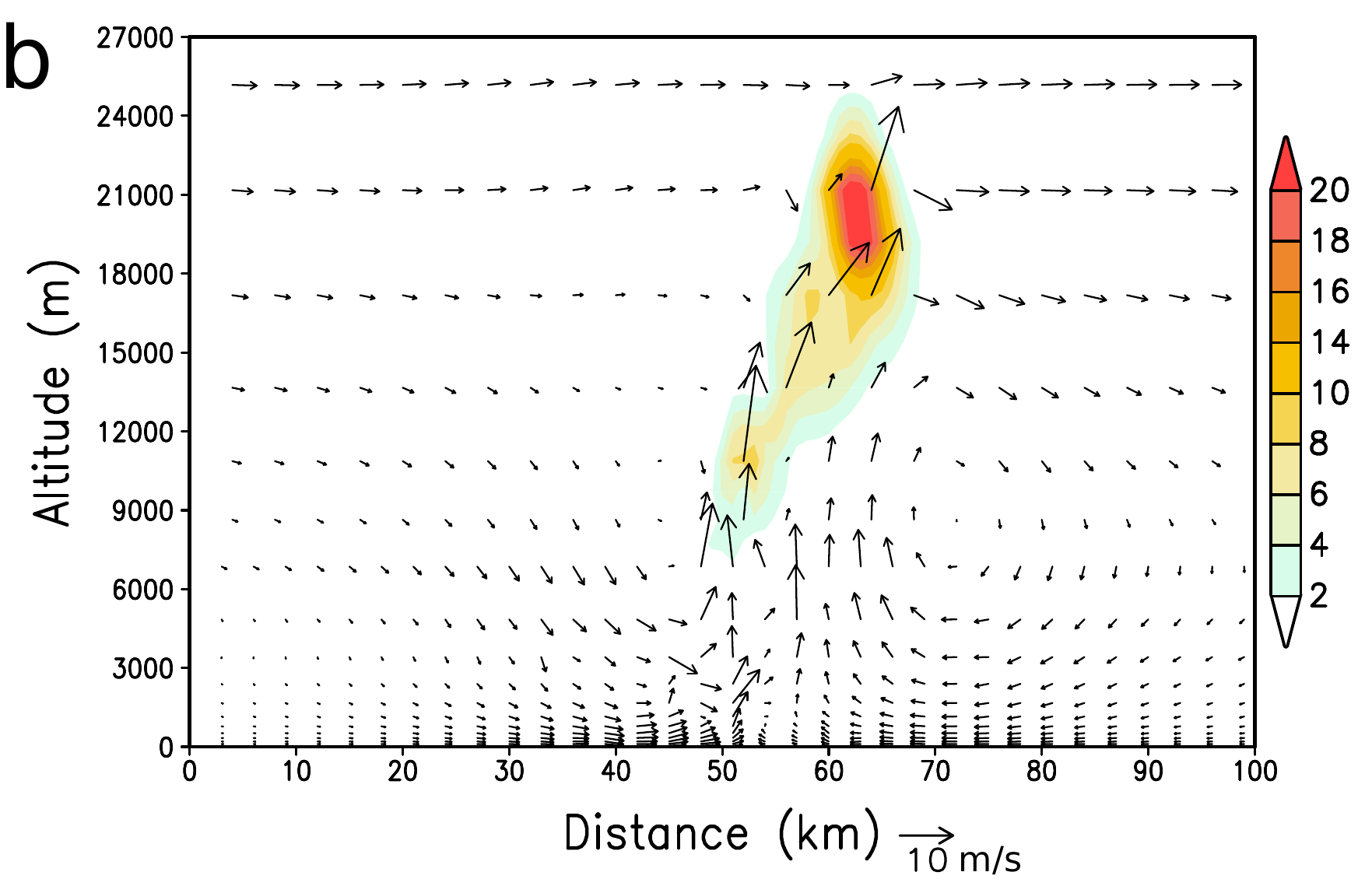}  
	\includegraphics[width=7cm]{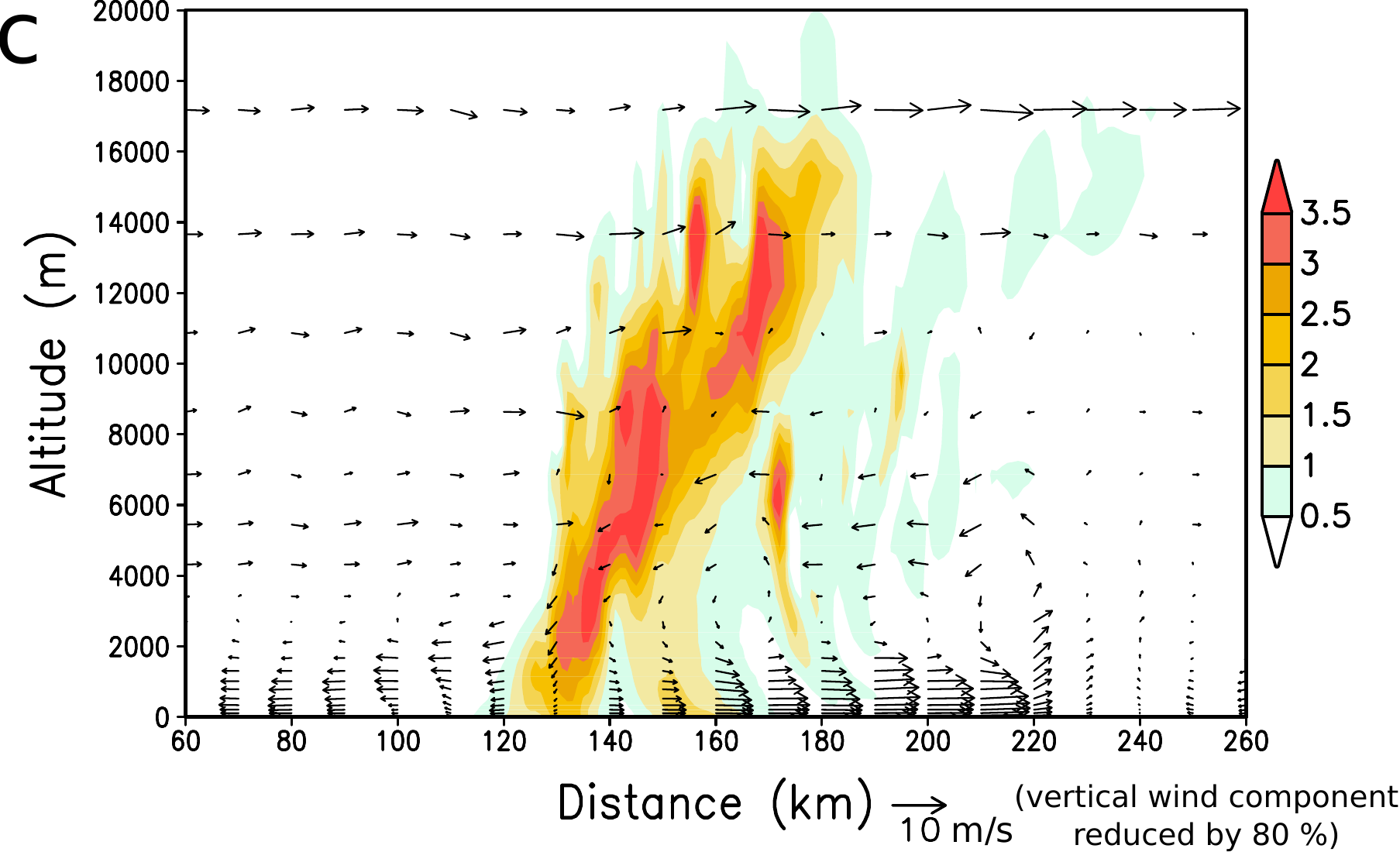}
        \includegraphics[width=7cm]{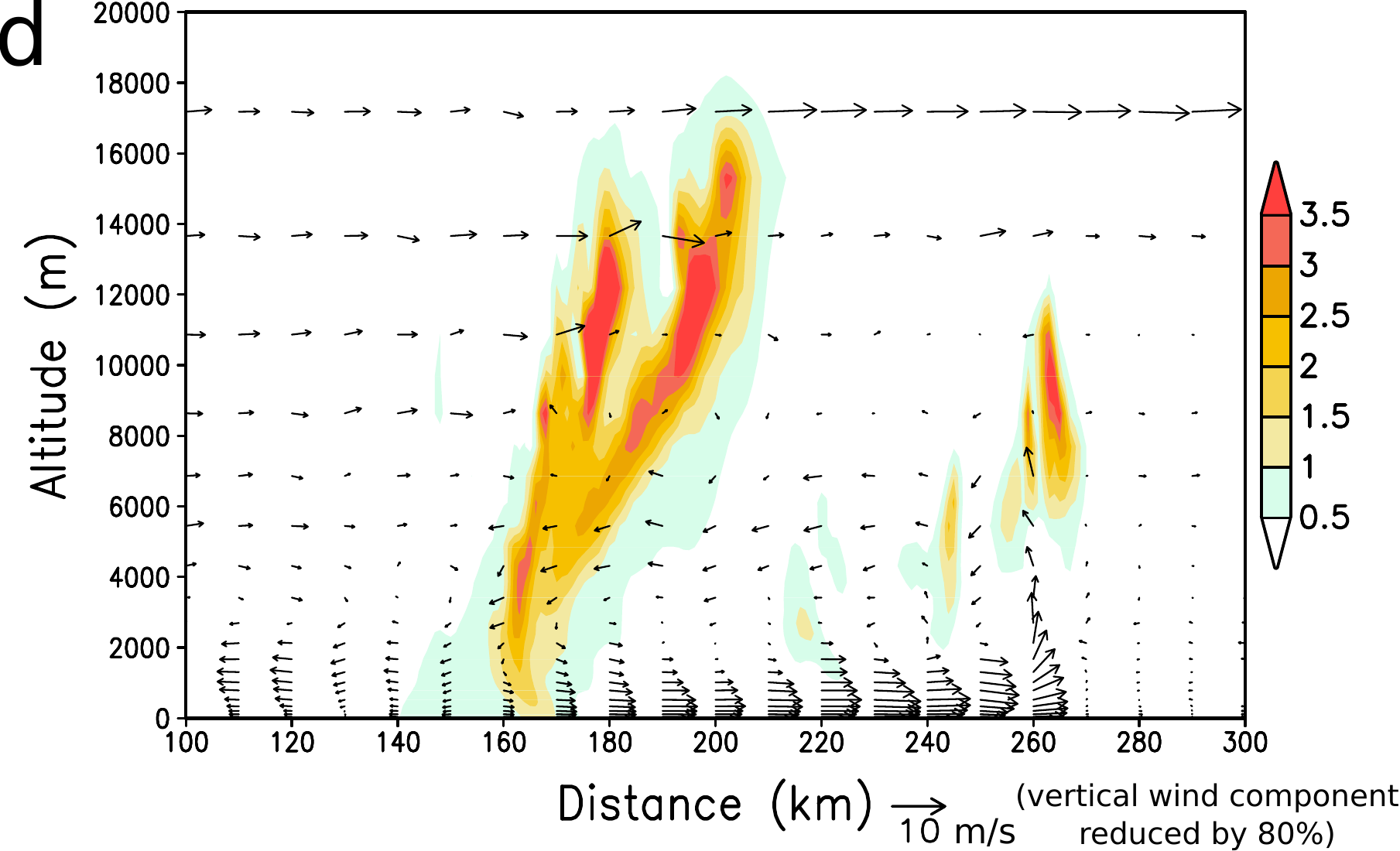}  
\end{center}  
\caption{\textbf{2D (altitude/longitude) simulation of the evolution of a storm under Titan's conditions at equator during equinox.} (a), (b), (c) and (d) are taken 1h, 1h40, 10h35 and 12h30 after the start of the simulation, respectively. 
The initial wind profile was derived from the GCM. The initial methane humidity corresponds to a CAPE of 500 J/kg. Color bars indicate the mixing ratio of condensed methane in g/kg. The wind vectors are scaled to the axis. A reference vector of 10 m/s for zonal wind is shown. In (c) and (d), the vertical wind component is reduced by 80 \% to better see the gust front.}
\label{figure_1}
\end{figure} 

For the initial conditions, we used the temperature profile measured by Huygens \citep{fulchignoni05} and a wind profile from {the Titan IPSL GCM with no methane cycle \citep{lebonnois12}. 
Huygens' measurements indicated a methane humidity of 40$\%$ at the surface \citep{niemann05}. These conditions do not allow the development of convective clouds but were measured at the end of the southern summer. For our simulations at the equinoctial season, we used a methane profile with a surface humidity of $\sim$50$\%$ or 60$\%$, which is necessary to form convective clouds and corresponds to a convective available potential energy (CAPE) of 250 or 500 J/kg.
The deep convection was triggered by a warm bubble of +2 K (i.e. the air temperature is increased in the first km). In our simulations, convective clouds reach an altitude of 25 km where flow fast eastward winds, and the storm dynamics leads to the formation of a eastward gust front above the surface. 
Fig. \ref{figure_1} shows the development of a gust front in one of our simulations. The triggering of the convection is associated with symmetric convergence of air below the cloud (Fig. \ref{figure_1}a). The cloud  develops obliquely because of the wind-shear (Fig. \ref{figure_1}b). 
To compensate for the ascending air in the storm, a downdraft forms behind the storm (Fig. \ref{figure_1}b). Passing below the cloud, it is cooled by rain evaporation, which accelerates it, and flows at the surface, producing the gust front.
Subsequently, the gust front ends up beyond the cloud (Fig. \ref{figure_1}c). It raises surrounding air from the surface to 3-4 km, producing secondary convective clouds that can increase its energy and life-time (Fig. \ref{figure_1}d). The passage of the gust front also produces a local peak of eastward surface wind (Fig. \ref{figure_2}a), traveling over large distances (in the order of 500 km in our simulations). Behind the gust front, wind speeds are typically one order of magnitude higher than usual winds close to the surface, and can reach 10 m/s during up to 9 hours (Fig. \ref{figure_2}a and b).

\begin{figure}[!h] 
\begin{center} 
	\includegraphics[width=7cm]{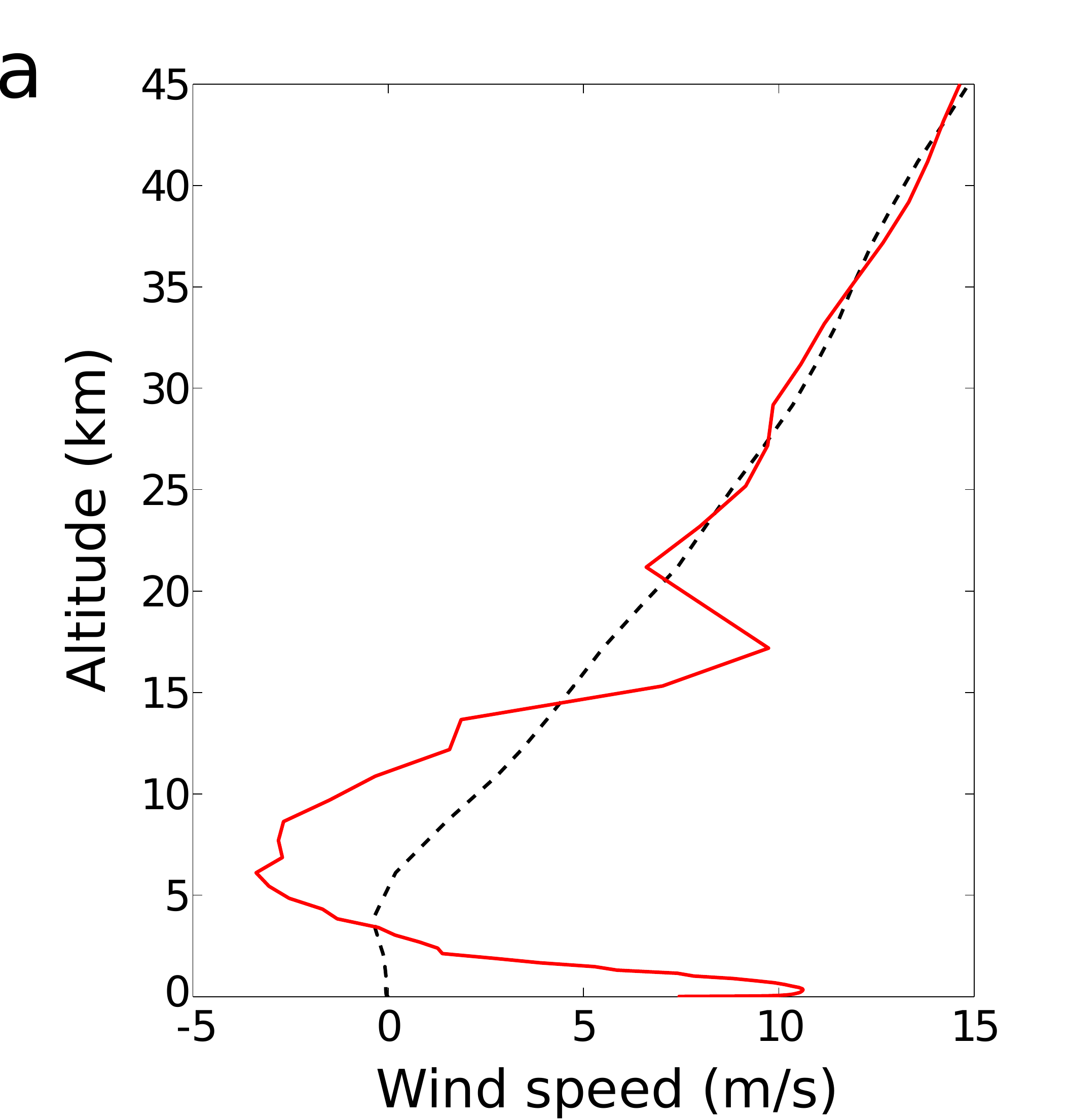}
 	\includegraphics[width=9cm]{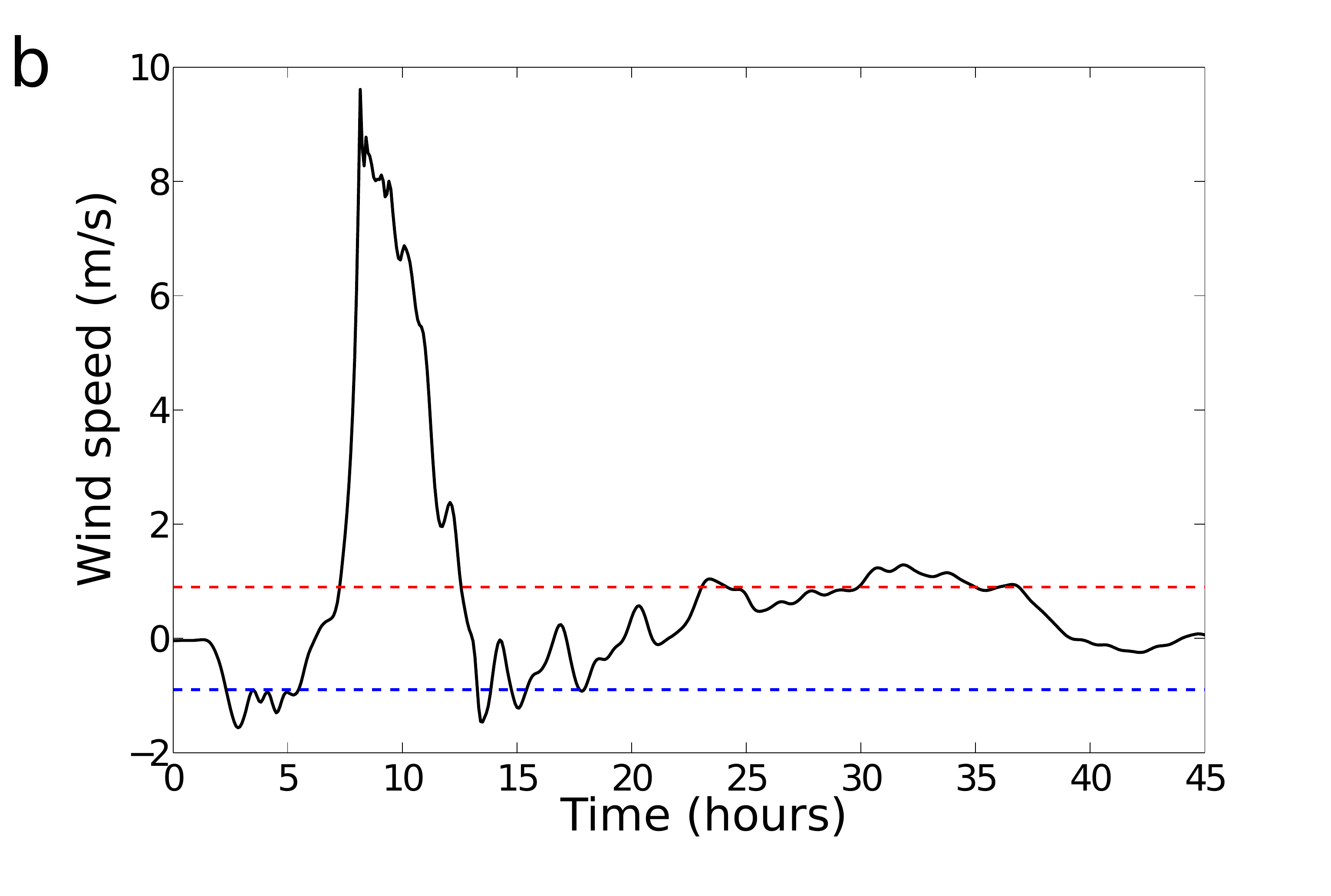}
\end{center} 
\caption{\textbf{Wind speed in the gust front.} (a) Wind profiles in the gust front (red) and without storm (black dashed line).
(b) Evolution of the zonal wind at 40 m (black line) during the passage of a storm. Eastward winds are positive. The red (blue) dashed line corresponds to the threshold friction speed of 0.04 m/s, i.e. 0.9 m/s at 40 m, for eastward (westward)} winds.
\label{figure_2}
\end{figure}

Our 2D simulations are representative of large storms or mesoscale convective systems. Individual small storms (e.g. less than 30 km in diameter) should produce weaker and more isotropic gusts than in our simulations. Yet, multiple small storms should interact with each other and organize the convection to form large storm or mesoscale convective systems, as the giant cloud systems observed in September and October 2010 \citep{turtle11b}, producing strong gust fronts with an eastward propagation. The dynamics of such large storms should be essentially 2D and qualitatively well represented by our idealized 2D simulations (see Supplementary Information).
We estimate that each region of the equator might statistically undergo on the order of one gust front every equinox (i.e. 2 per Titan year, see Supplementary Information).

The nature of the sand on Titan is not well known. The optimum grain size for saltation is estimated to be around 300 $\mu$m with a threshold friction speed around 0.04 m/s \citep{lorenz13}, corresponding to a wind speed around 0.9 m/s  at 35 m above the surface (see Supplementary Information). 
In our GCM, winds at 35 m and at the equator are generally below 0.5 m/s and do not exceed 1 m/s (see Supplementary  Fig. 6 and Fig. 7). Thus, usual winds on Titan are likely to generate only limited sand transport.  
Our GCM winds follow a Weibull distribution. They exceed the threshold friction speed of 0.04 m/s only 0.06$\%$ of the time, producing a westward sand flux at the equator of only 0.0015 m$^2$/yr (see Supplementary Information for the sand flux calculation). However, GCMs tend to underestimate sand flux as they miss local gusts \citep{bridges12}. To compensate for this, we also consider a 10 times stronger westward sand transport (0.018 m$^2$/yr) by increasing GCM wind speed by 20$\%$.
Storm gusts largely exceed the wind threshold by an order of magnitude. According to our mesoscale simulations, the sand flux produced by storms is 2.7 times stronger eastward than westward, and one storm per equinox (i.e. two per Titan year) produces on average an eastward sand flux around 0.15 m$^2$/yr. Although episodic, tropical storms should therefore dominate the sand transport. A higher saltation threshold friction speed (e.g. 0.052 m/s)  \citep{burr15} would strongly reduce the sand flux from general circulation winds, while weakly impacting sand transport by storm gusts.

On Earth, while in unlimited sediment  many linear dunes are oriented orthogonally to the direction of maximum sediment transport \citep{rubin87}, in limited sediment they align with resultant transport direction, called the resultant drift direction (RDD), and propagate over long distances with regular spacing and a limited amount of sediment in interdune areas \citep{courrech14}. This mechanism may explain the formation and the evolution of Titan's dunefields. 
It requires neither cohesion \citep{rubin09} nor complex secondary flows, but relies on the presence of sedimentary reservoirs and multidirectional wind regimes producing an overall sediment flux over a non-erodible ground \citep{courrech14}. Given that many locations in Titan's dune fields appear to have sand-free bedrock interdunes \citep{rodriguez14}, the conditions of linear dune down-axis extension seem to be met.

Fig. \ref{figure_3}a shows the RDD combining winds from the Titan IPSL GCM with no topography and storm gusts from mesoscale simulations, with respect to the storm frequency. With no storm or a very small frequency of storm, the RDD  is oriented westward with a southward component due to Saturn's eccentricity (the latter results in stronger solar forcing and increased Hadley circulation during the southern summer, hence producing southward winds stronger than northward winds during the northern summer). With around two storm per Titan year (i.e. one per equinox), the sand tranport produced by equinoctial storms dominates and the RDD is oriented eastward. With such a frequency, the storm control remains efficient even for the GCM winds with increased gust.

\begin{figure}[!h] 
\begin{minipage}[c]{.5\linewidth}
\begin{center}
       \includegraphics[width=8cm]{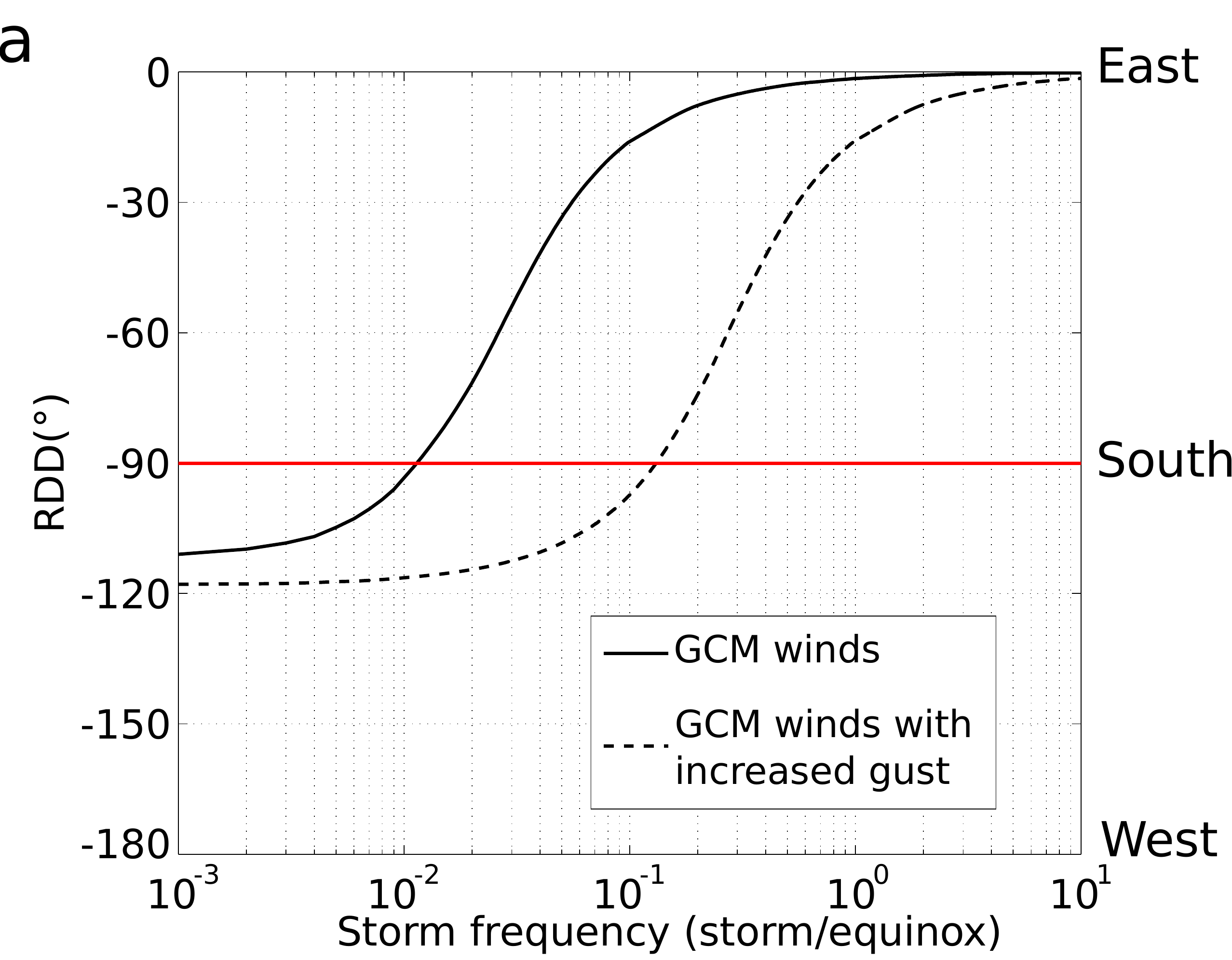}
\end{center} 
\end{minipage} 
\begin{minipage}[c]{.35\linewidth}
\begin{center} 
 	\includegraphics[width=8cm]{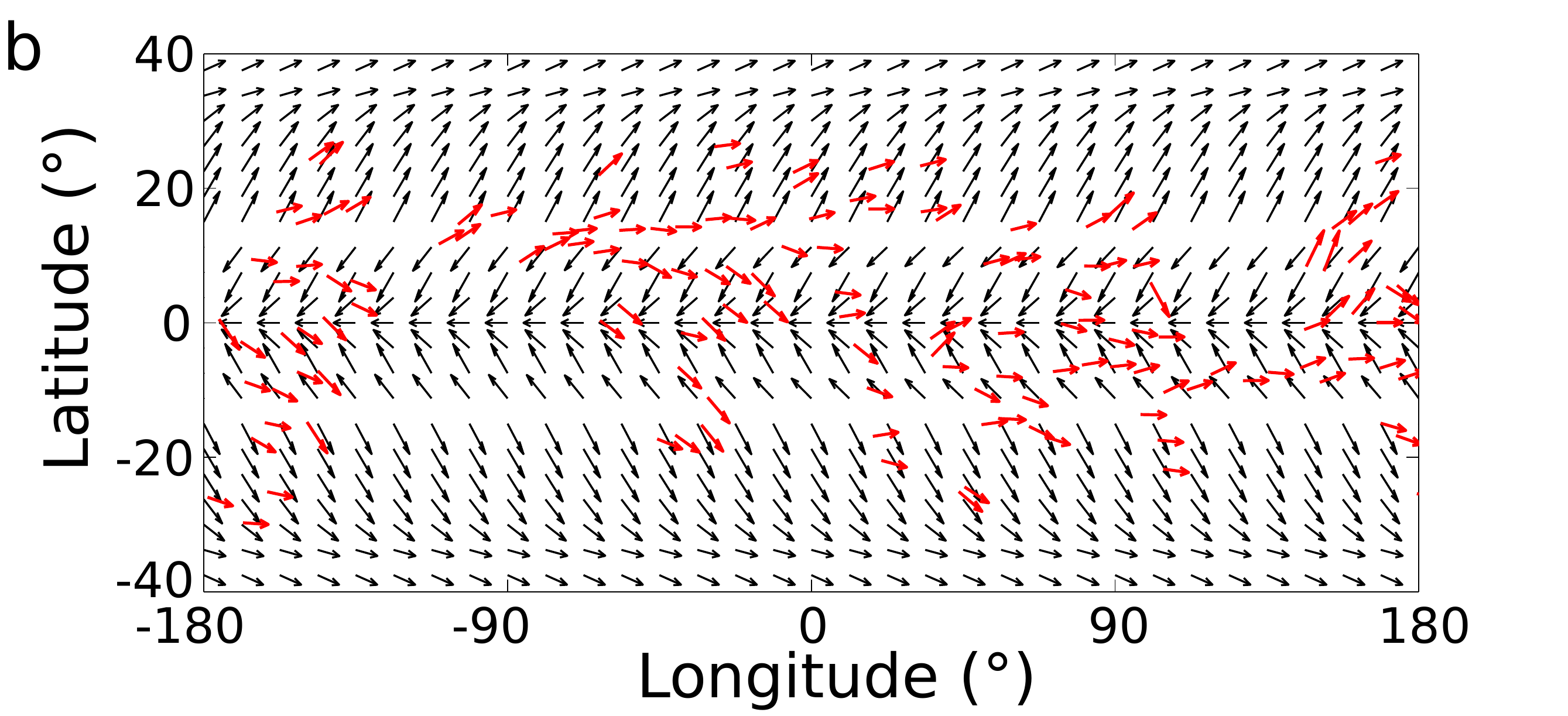}
	\includegraphics[width=8cm]{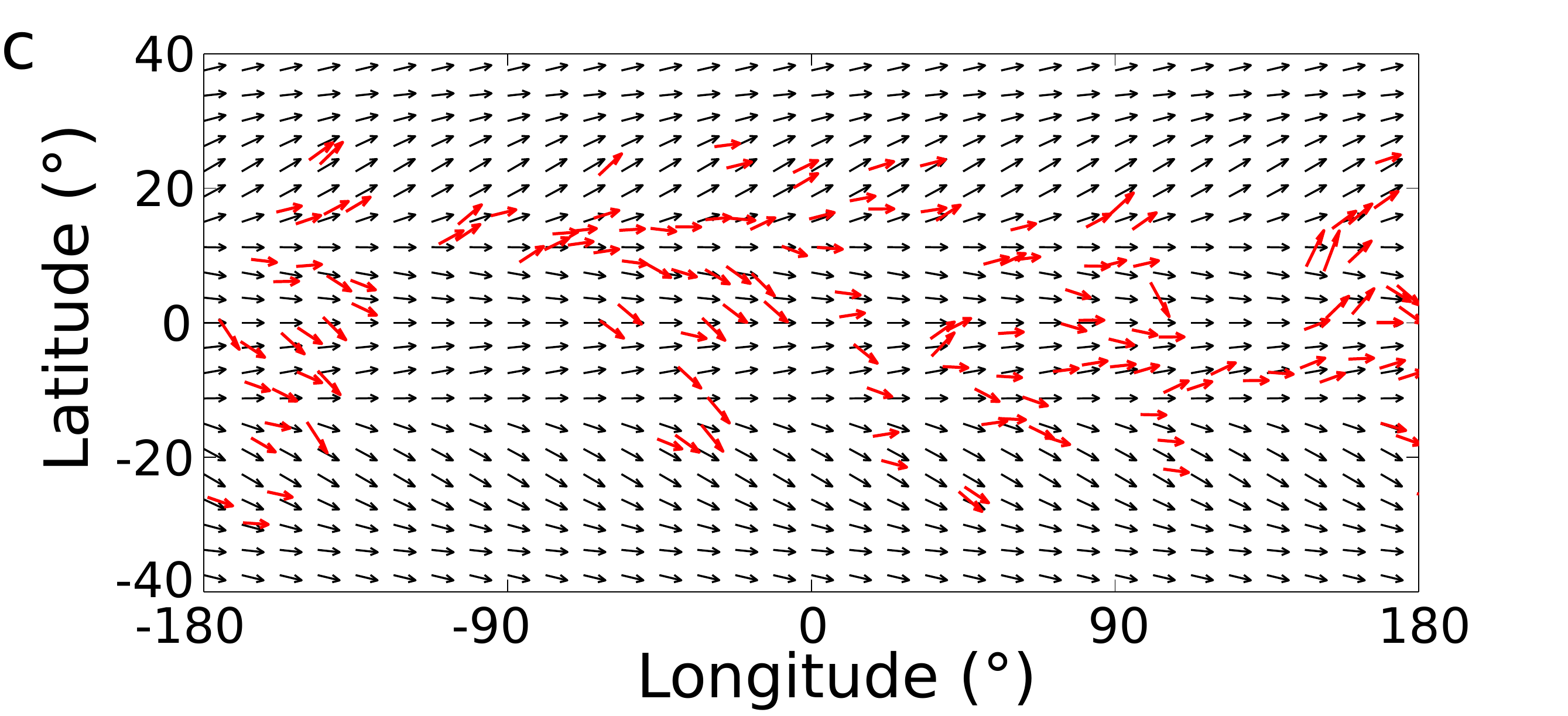}
\end{center} 
\end{minipage}

\caption{\textbf{Storm impact on the resultant drift direction (RDD).} 
(a) The RDD as a function of the storm frequency per equinox with a saltation threshold of 0.04 m/s, using GCM winds (black solid line) or GCM winds with increased gust (black dashed line). Angles are measured anti-clockwise from the East (i.e. -180$^\circ$ is westward, -90$^\circ$ is southward and 0$^\circ$ is eastward). The red line symbolizes the passage from westward to eastward dune growth.
(b) Map dune orientation observed with Cassini's radar \citep{lucas14} (red) and map of RDD obtained with the GCM (with increased gust) for low latitudes with a threshold of 0.04 m/s and no storm effect (black). (c) is same as (b) but with the impact of two storms per Titan year.} 
\label{figure_3}
\end{figure}

For a 100 m-high dune, the eastward sand flux produced by two storms per Titan year (i.e. 0.15 m$^2$/yr) corresponds to an extension rate of 3 mm/yr (see Supplementary Information). According to the typical length of Titan's dunes (i.e. 30-50 km) \citep{radebaugh08}, this implies a minimum formation time of around 15 Myr. This time is longer than the period of Saturn's perihelion precession (i.e. 45 kyr), which corresponds to the switch of the warmest summer between both hemispheres. Thus the southward component due to this orbital effect should vanish in dune orientation. Barchans and small dunes have been observed with shifted orientation as compared to long dunes \citep{ewing15}. This has been interpreted as the effect of a long-term, cyclic multimodal wind regime. While small dunes may respond to long orbital changes (affecting both general circulation winds and storm impact), our multimodal wind pattern can also explain their shifted orientation by growth orthogonal to the direction of maximum sediment transport \citep{rubin87, courrech14}.
Fig. \ref{figure_3}b and Fig. \ref{figure_3}c show the RDD map in the equatorial band  without or with the effect of two storms per Titan year. 
In both cases, we averaged the effect of Saturn's eccentricity according to the above discussion.
Our GCM predicts a RDD slightly converging to the equator for latitudes lower than 12$^\circ$ N/S and diverging from the equator for latitudes higher than 12$^\circ$ N/S. The divergence is due to faster poleward summer winds. This may explain the observed poleward orientation of dunes for latitudes higher than around 10$^\circ$ N/S \citep{lorenz09, lucas14}. 
These poleward sand fluxes also imply that Titan's sand cannot be transported from polar regions to the equatorial band. The sand of Titan's dunes must therefore have been produced in equatorial regions.

The sand flux pattern that we estimate for Titan's dunes is similar to those implicated in the formation of longitudinal dunes in the Rub'al-Kali desert (southern Arabia) and in the Great Sand Sea (Egypt). In these deserts subject to a dominant wind, linear dunes are parallel to the RDD (see Fig. \ref{figure_4}b and Supplementary Fig. 10). We can therefore establish an analogy with Titan's dune formation (see Fig. \ref{figure_4}).
In terrestrial deserts, gust fronts formed by convective clouds can produce dust storms, also called haboobs \citep{miller08}. Similarly, gust fronts generated by Titan's storms could produce dust storms. Indeed, their friction speed might exceed the threshold for micrometric particles (around 4 m/s at 35 m for a 10 $\mu$m diameter particles, see Supplementary Information). 
They could also transport centimetric pebbles (threshold speed around 4-9 m/s at 35 m for 1-5 cm diameter pebbles), explaining the diffuse material trailing out to the east around many mountains \citep{radebaugh08}.

 \begin{figure}
\begin{center} 
 	\includegraphics[width=9cm]{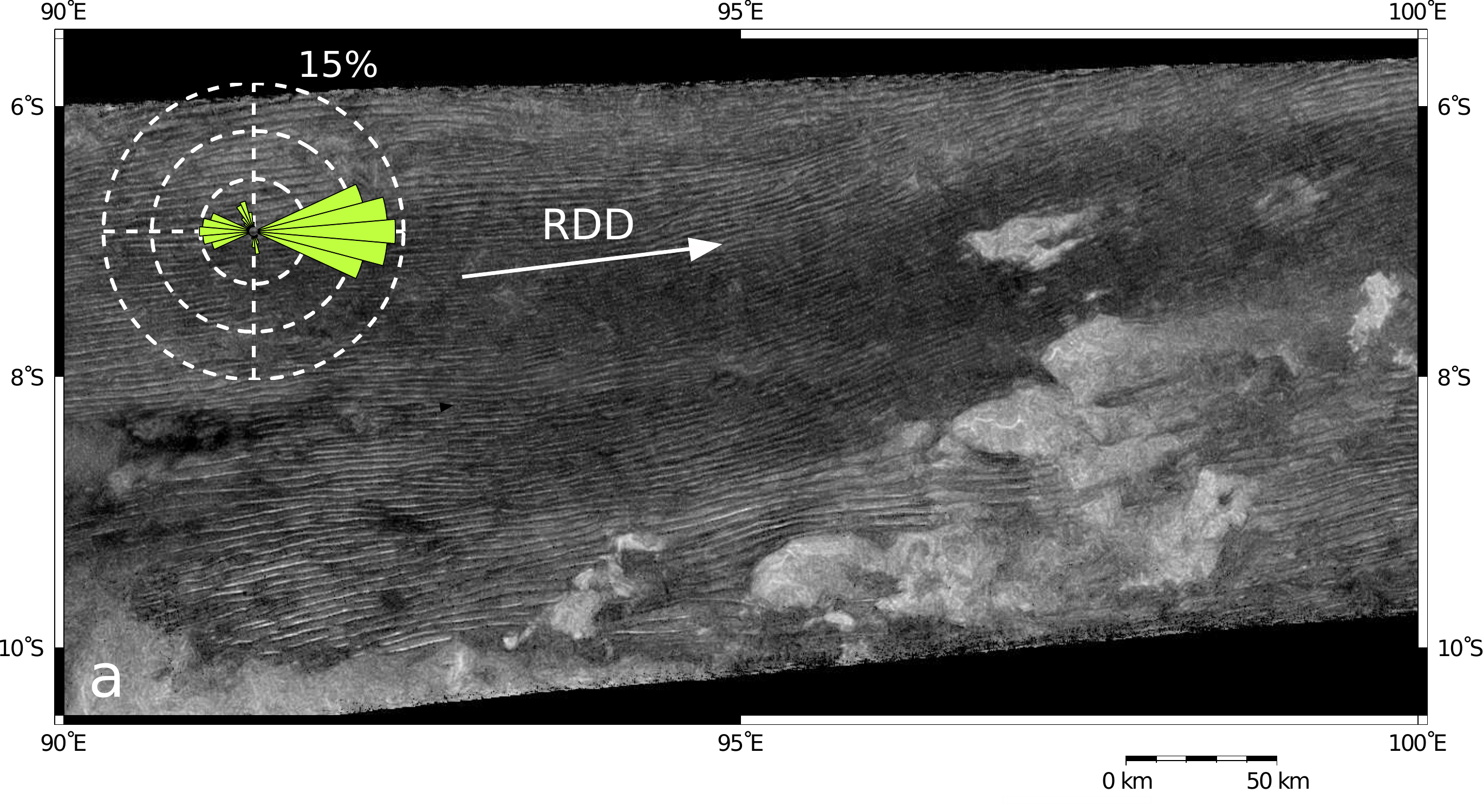}
 	\includegraphics[width=7cm]{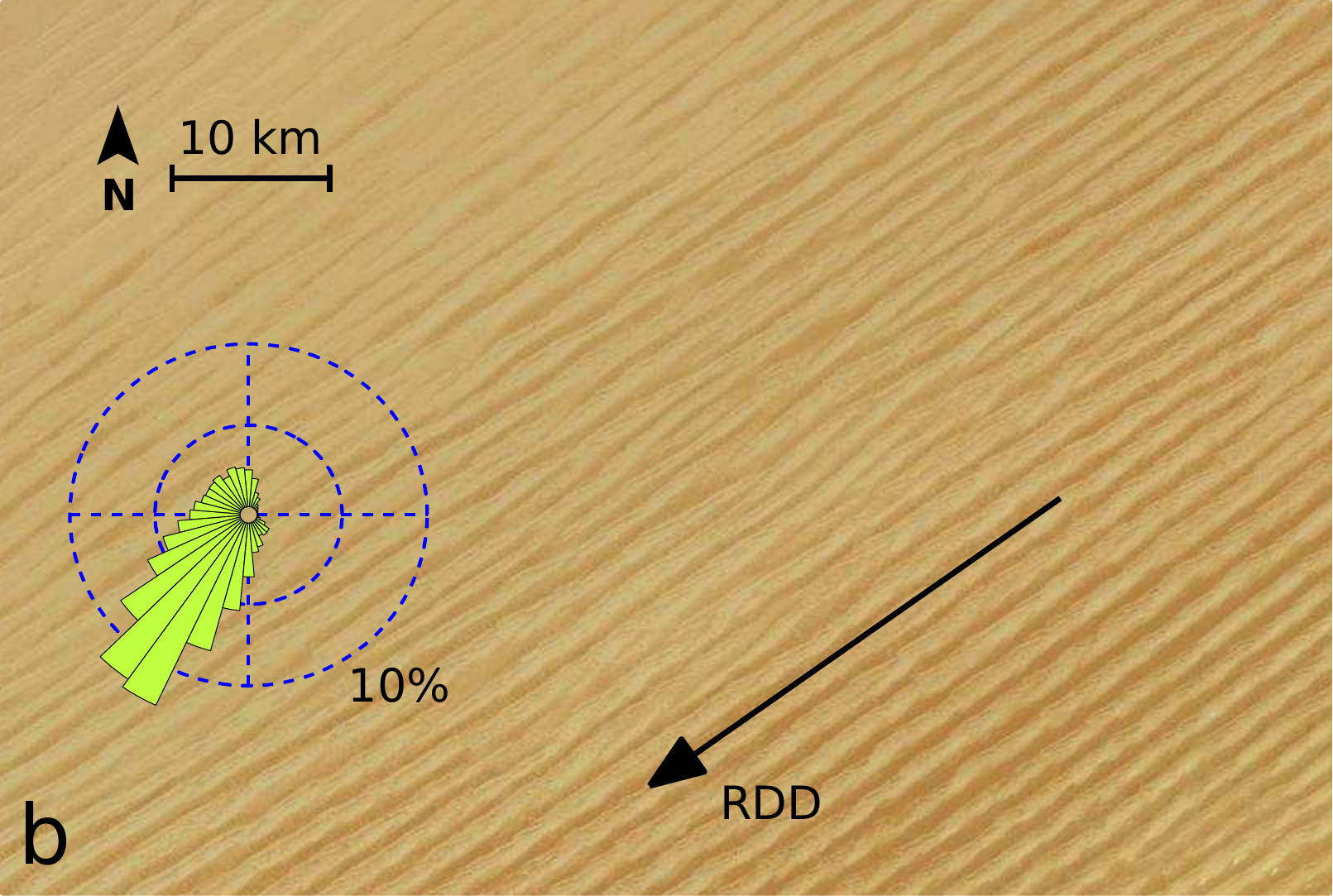}
\end{center} 
\caption{\textbf{Analogy between linear dunes on Titan and in Rub'al-Kali desert on Earth.} (a) denoised image of Titan's dunes from Cassini's radar SAR (see Supplementary Information). The inset shows the sand flux rose similar to Fig. \ref{figure_3}b, but calculated at 7.5$^\circ$S and averaging the effect of Saturn's eccentricity.
(b) longitudinal dunes in Rub'al-Kali desert (18$^\circ$ N, 48$^\circ$ E) with the sand flux roses calculated from winds at 10 m. For both images, the arrow corresponds to the resultant drift direction calculated with the sand flux rose.} 
\label{figure_4}
\end{figure}

In conclusion, the combination of GCM and mesoscale simulations, together with analogies with terrestrial dune formation, provides a comprehensive scheme to explain the major features of Titan's dunes (i.e. eastward propagation, linear shape, poleward divergence and 2-3 km spacing). 
This analysis highlights the potential importance of storms and mesoscale phenomena on aeolian processes. Such rare and strong events may play or have played a important role in some dune fields on Earth and Mars.

\paragraph{}
\textbf{\large Methods}

Our analysis was performed using GCM and mesoscale simulations, together with analogies with terrestrial dunes. 
We use the Titan IPSL GCM \citep{lebonnois12,charnay12} to predict surface winds in the equatorial regions. This GCM reproduces well the superrotation and the thermal structure measured by the Huygens probe in the troposphere. Simulations are performed taking into account the effects of the diurnal cycle and Saturn's gravitational tides, but with no topography and no methane cycle (see also Supplementary Information). Titan's climate is essentially dry, excepted in polar regions. The methane cycle is therefore not expected to significantly affect the tropospheric dynamics in the equatorial regions. This is illustrated by the excellent match between the thermal structure and its different layers measured by the Huygens probe and simulated by the GCM \citep{charnay12}. To compute wind statistics and sand transport, we use instantaneous GCM surface winds over a year and all longitudes.

Methane storms are simulated with the mesoscale model TRAMS in 2D including  full methane cloud microphysics (see also Supplementary Information) \citep{barth07,barth10}. Incidentally, a GCM including the methane cycle cannot resolve storm cold pools and gust fronts, which are subgrid phenomena. As boundary conditions, we use the GCM zonal wind profile at the equator and at the north spring equinox, the Huygens' temperature profile \citep{fulchignoni05} and the Huygens' methane humidity profile with an increase in the lower troposphere. These idealized simulations are representative of large or squall line storms.
We estimate the order of magnitude of the storm frequency at each location in the equatorial band around one storm every equinox. A simple estimation of the size and displacement of the giant arrow shape storm observed by Cassini at the equinox in 2009 \citep{turtle11a,turtle11b} gives a lower limit of 0.2 storm per equinox (see Supplementary Information).
To compute the sand transport and sand flux roses produced by general circulation winds and gust fronts, we use a sand flux formula with a saltation threshold speed estimated from previous studies \citep{lorenz13}.
We also establish an analogy between Titan's dunes and some terrestrial dune fields, in particular in southern Arabia and Egypt (see Fig. \ref{figure_4} and Supplementary Fig. 10), where we predict sand fluxes from wind reanalysis data.
The simulation data are available from B.C. on request.

\paragraph{}
 
\paragraph{Acknowledgements} We thank J.-Y. Grandpeix, F. Forget, A. Spiga and J. Leconte for helpful discussions. 
We acknowledge financial support from  the UnivEarthS LabEx program of Sorbonne Paris Cit\'e
(ANR-10-LABX-0023 and ANR-11-IDEX-0005-02), the French National Research Agency (ANR-12-BS05-001-03/EXO-DUNES) and the Centre National d'Etudes Spatiales.
B.C. acknowledges support from an appointment to the NASA Postdoctoral Program at NAI Virtual Planetary Laboratory, administered by Oak Ridge Affiliated Universities.

\paragraph{Author Contributions} B. C. developed the idea of the methane storm control. E. B. and S. R. developed  and ran the mesoscale model. S. L. and B. C. developed and ran the GCM. B. C. analysed the simulations.  C. N. and S. C. P. provided the dune growth mechanism and the figure 4b. A. L. provided the denoised image and orientations of Titan's dunes. B. C. wrote the paper with significant contributions from all the authors in interpreting the results and editing of the manuscript.

\paragraph{Author Information} Competing financial interests: The authors declare no competing financial interests. 
\\
Correspondence and requests for materials should be addressed to B. C. (benjamin.charnay@lmd.jussieu.fr).


\newpage

\begin{center} 
\LARGE{Methane storms as a driver of Titan's dune orientation}\\
\textit{Supplementary Information}
\end{center} 

 \section{Orientation of Titan's dunes from radar images}

\begin{figure}[!h] 
\begin{center} 
	\includegraphics[width=15cm]{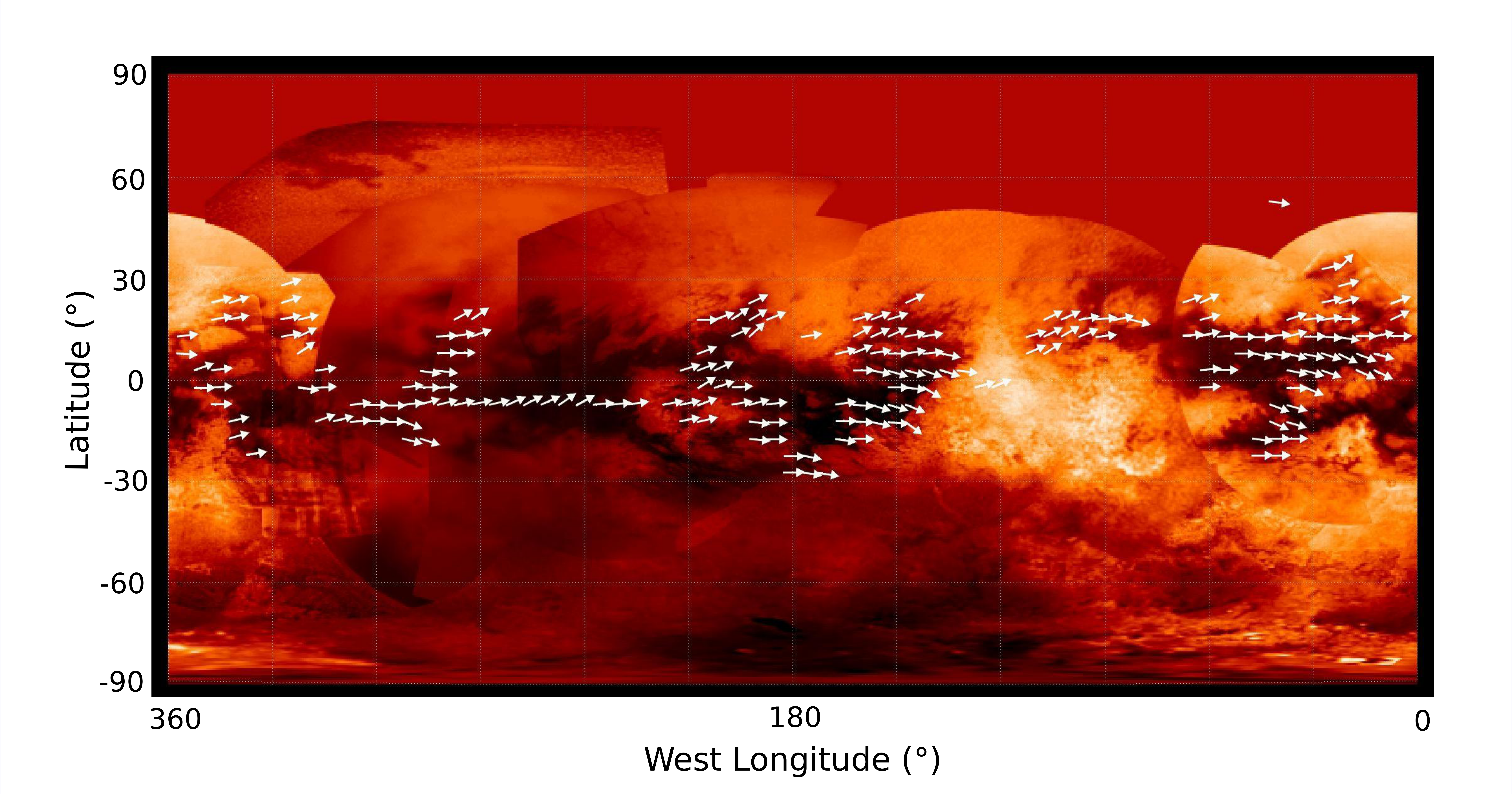}
\end{center} 
\caption{\textbf{Map of Titan's dune orientation.} Map of radar-measured dune orientation vectors \citep{lorenz09} (CREDITS: NASA/JPL-Caltech/ASI/Space Science Institute, PIA11801), showing the global eastward propagation and the divergence from the equator for latitudes higher than 10$^\circ$.}
\label{figure_s5}
\end{figure} 

Fig. \ref{figure_s5} shows the map of dune orientation and propagation, indicated with vectors, over a near-infrared basemap derived from Cassini ISS (Imaging Science Subsystem) images \citep{lorenz09}. The direction of propagation is obtained by looking at dune morphology around obstacles and dune terminations, with for instance some dunes stopping on the west side of obstacles and others diverting and recombining beyond the east side of obstacles. This analysis reveales that all dunes propagate eastward. Concerning the North-South orientation, a statistical analysis indicates that dunes tend to diverge poleward for latitudes higher than around 10$^\circ$ N/S\citep{lucas14}.

 Images of Titan's surface obtained by the Cassini's radar SAR (Synthetic Aperture Radar) suffer from errors from a variety of sources, the most prominent being speckle noise. This noise refers to the constructive and destructive interference of scattered energy from roughness elements on a scale smaller than the size of a SAR pixel. The result is a multiplicative noise, which hinders interpretation. An advanced denoising algorithm has been adapted to Cassini SAR data \citep{lucas14b}. It has been applied to the original image (i.e. the image by Cassini Radar, T8 flyby) of Fig. 4a of the letter to significantly reduce the noise.

Fig. 3a and 3b of the article show dune orientations obtained in the same way as Fig. \ref{figure_s5} but with denoised radar images whose dunes segments have been detected \citep{lucas14} and their orientation averaged over areas of 8$^\circ$$\times$8$^\circ$ in longitude-latitude.

 \section{Calculation of general circulation winds}
 \textbf{Description of the IPSL Titan GCM}\\
For this study, we used a 3-dimensional GCM \citep{lebonnois12,charnay12}. A horizontal resolution of 32 $\times$ 48, corresponding to resolutions of 3.75$^{\circ}$ latitude by 11.25$^{\circ}$ longitude, is used for the simulations.  This GCM covers altitudes from the ground (first level at 35 m) to 500~km. The dynamical core is based on the most up-to-date version of the LMDZ \citep{hourdin06}. It is a finite-difference discretization scheme that conserves both potential enstrophy for barotropic nondivergent flows, and total angular momentum for axisymmetric flows. The version used in this study includes gravitational tides \citep{tokano06b}, though the impact in the troposphere does not influence the effects described in this work. We found tidal effects on the pressure similar to previous works \citep{tokano06b}, but tidal winds are much weaker in our model. We use a fully coupled aerosol microphysics calculated in 2D (zonally averaged)  \citep{rannou04b}. The present model is dry and does not take into account the methane cycle. The profile of methane is fixed (close to the HASI profile \citep{niemann05}) for the radiative transfer. The latter is based on the McKay radiative code \citep{mckay89} and includes the diurnal cycle. The radiative transfer  is called 200 times per Titan day.
For the surface, we use an albedo of 0.3, a rugosity length of 0.005~m, an emissivity of 0.95 and a thermal inertia of 400~J/m$^{2}$/K. In this study, we run the GCM with a flat topography.

To calculate the friction speed $u_{\star}$  from the GCM wind $u$ at an altitude $z$, we use the relation:
\begin{equation}
u_{\ast}=\frac{\kappa}{\ln(z/z_0)} u
\end{equation}
with $\kappa$ = 0.4 the Von-Karman constant and $z_0$ = 0.005 m the rugosity length.
The threshold $u_{\ast t}$=0.04 m/s corresponds to a wind speed of 0.89 m/s at 35 m.
\\
\\
\textbf{GCM wind statistics:}\\
For all calculation implying the GCM wind statistics (e.g. sand fluxes), we use instantaneous GCM winds with 20 outputs per Titan day and combining wind series at all longitudes. The dissimilarities for wind statistics obtained for different Titan years are negligeable.
Fig. \ref{figure_s1} shows the surface wind roses produced by the GCM in the equatorial band.
Surface winds are essentially bimodal. At the equator, they blow from NE to SW in northern winter and from SE to NW in northern summer. For latitudes higher than 5$^\circ$ N/S, the summer winds are eastward. Because of Saturn's eccentricity, the southern summer is shorter and hotter than the northern summer. This implies that northerly winds are less frequent but stronger than southerly winds.
\\
\begin{figure}[!h] 
\begin{center} 
	\includegraphics[width=5.6cm]{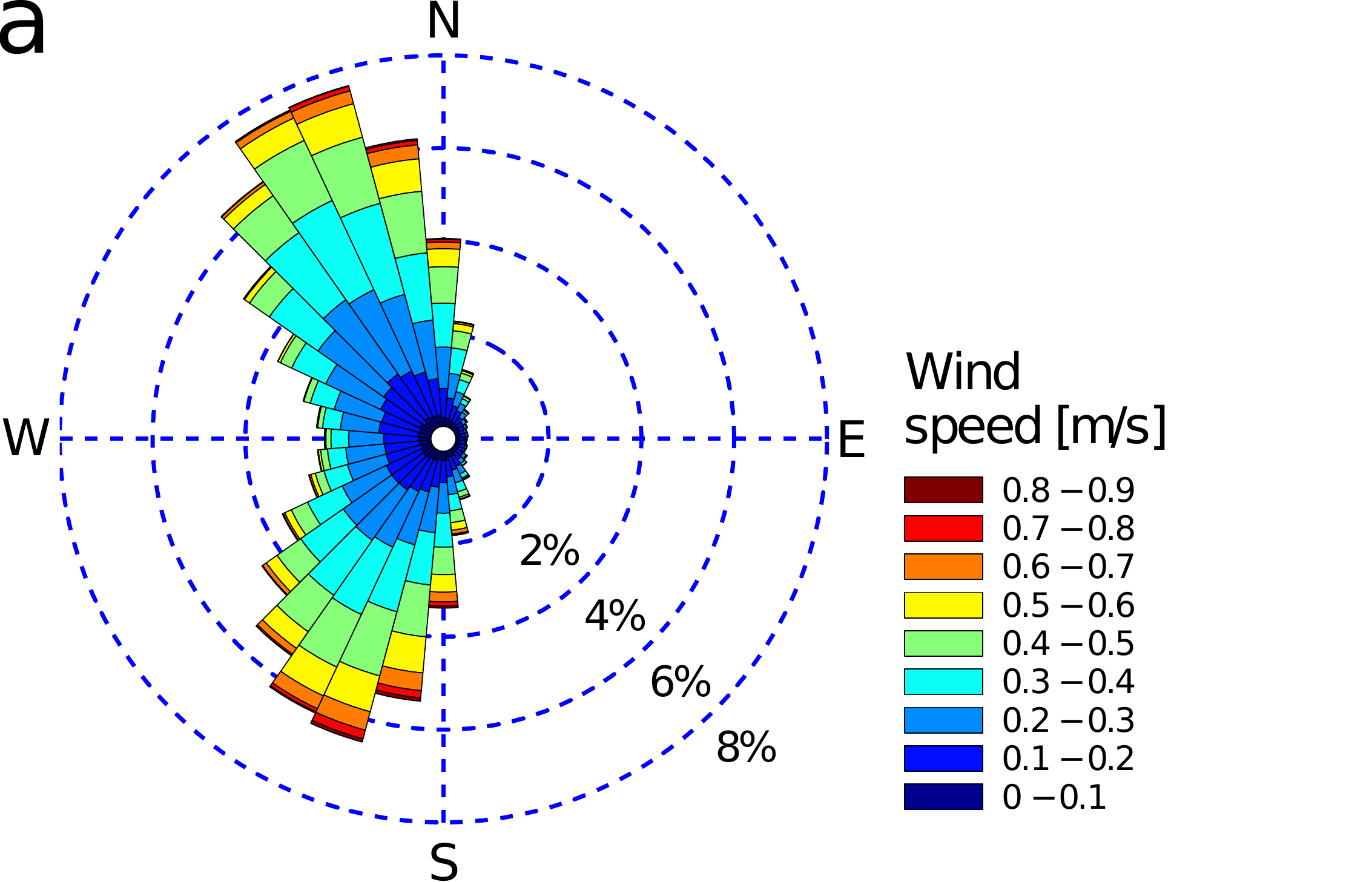}
       \includegraphics[width=5.6cm]{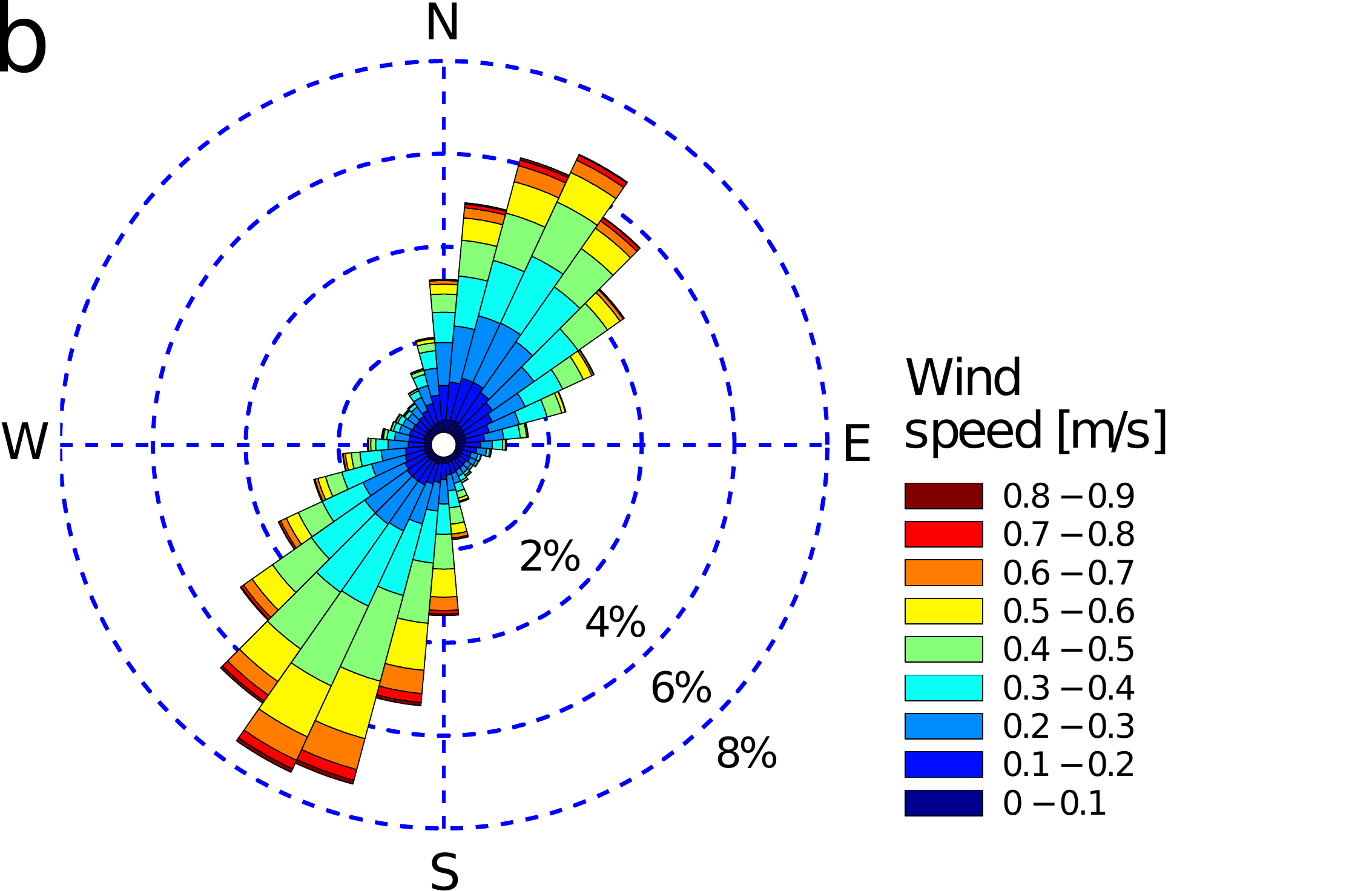}
 	\includegraphics[width=5.6cm]{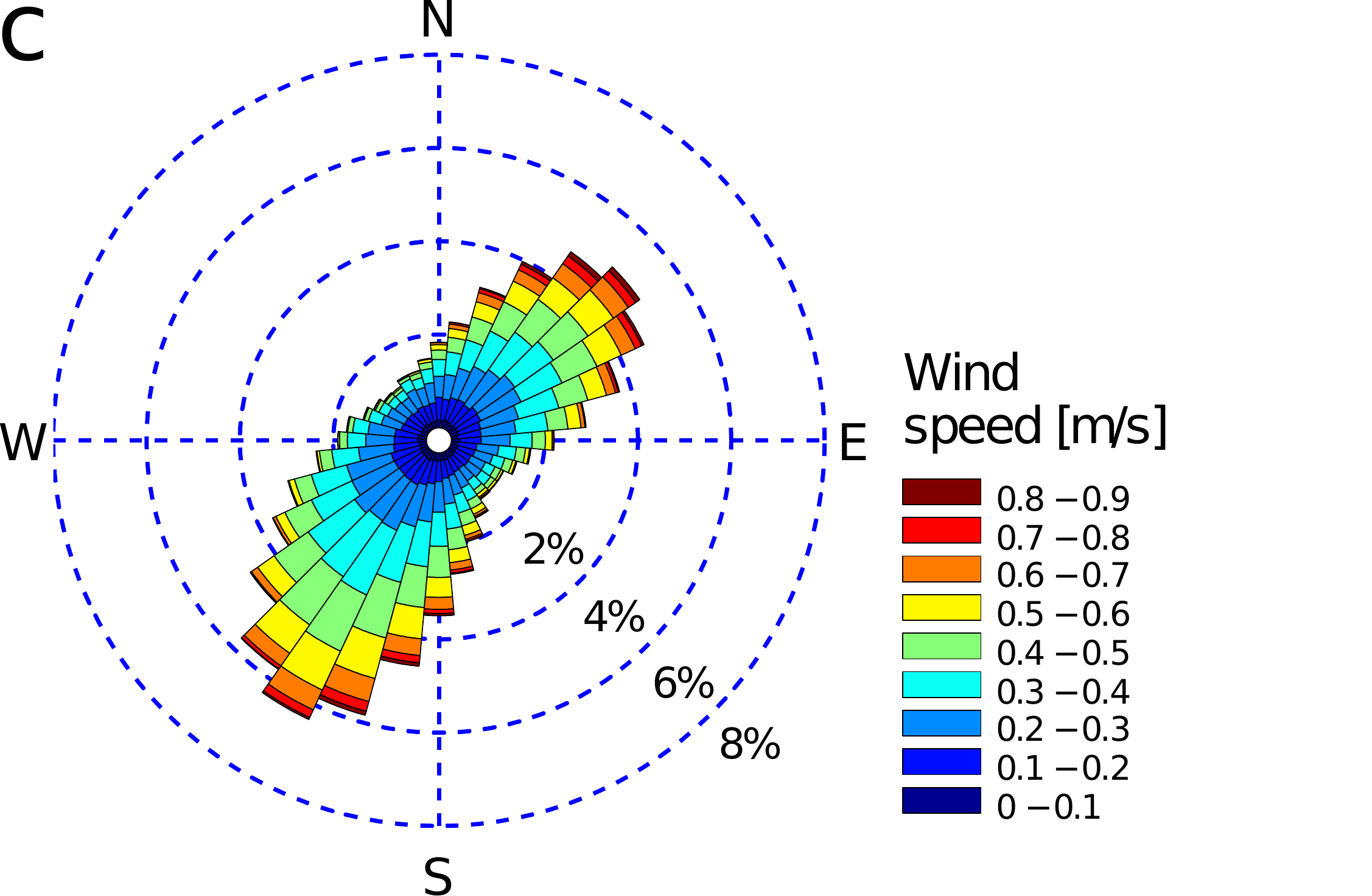}
\end{center} 
\caption{\textbf{Wind roses (direction and speed) produced by the GCM}. The roses have been obtained with instantaneous winds at 35 m at 0$^\circ$N (left) 10$^\circ$N (middle) and 20$^\circ$N (right). The frequency is given in percent.}
\label{figure_s1}
\end{figure} 
\\
\begin{figure}[!h] 
\begin{center} 
	\includegraphics[width=9.cm]{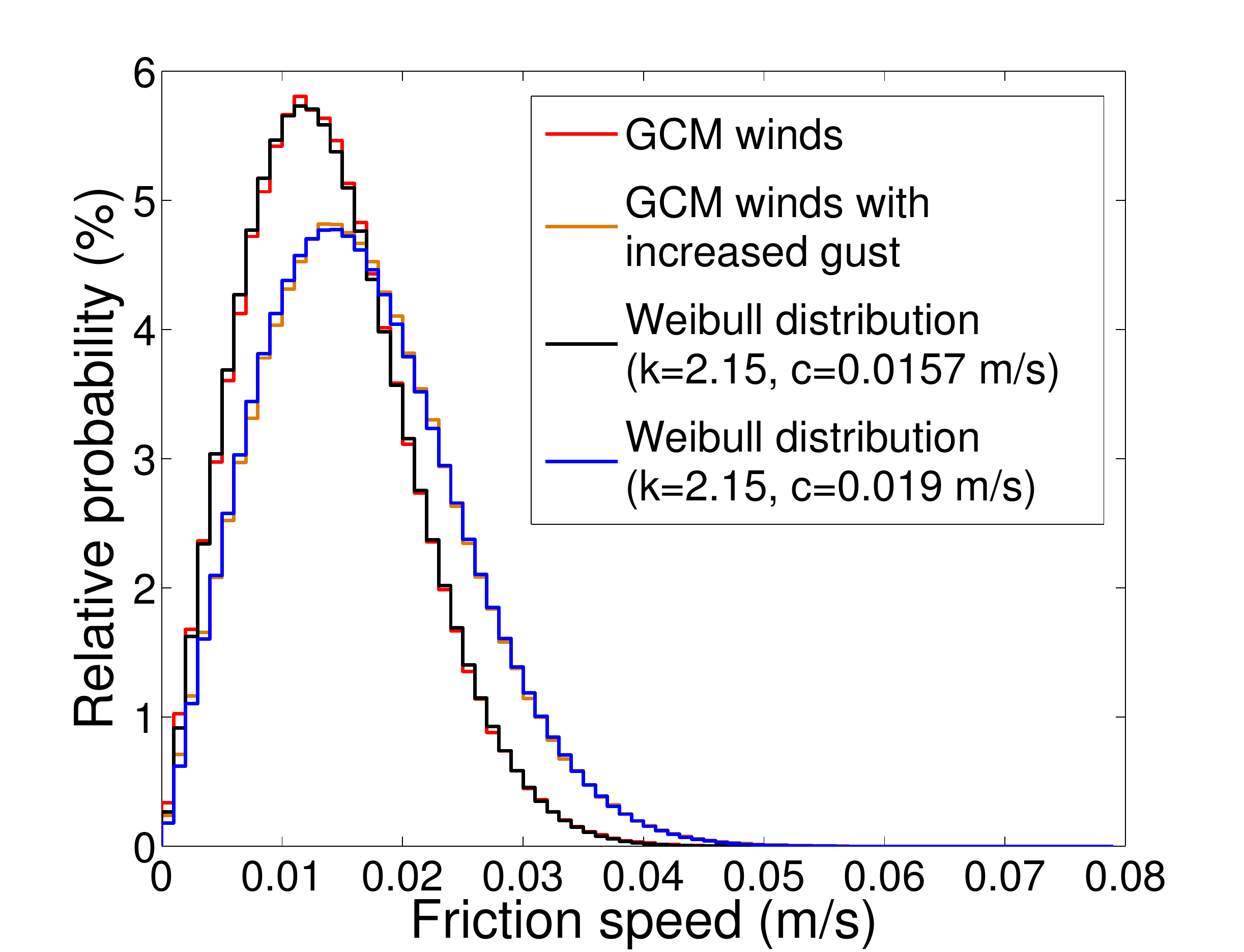}
\end{center} 
\caption{\textbf{Statistics of GCM winds at the equator.} The different lines correspond to the relative probability of friction wind speed per bin of 0.001 m/s for the GCM winds (in red) and the GCM winds with increased gust (in orange) at the equator. The black line corresponds to the Weibull distribution fitting the GCM wind statistics (coefficient k=2.15 and c=0.0157 m/s) and the blue line the distribution fitting the GCM wind statistics with increased gust (coefficient k=2.15 and c=0.019 m/s).}
\label{figure_s2}
\end{figure} 

Fig. \ref{figure_s2} shows the relative probability of the friction speed from the GCM (in red) at the equator and per bin of 0.001 m/s.  Wind speed exceeds the threshold friction speed of 0.04 m/s only around 0.06 $\%$ of the time. The GCM wind statistics are well described by a Weibull distribution (in black in the figure), for which the probability of exceeding a friction speed $U$ is $P(>U)=\exp(-(U/c)^k)$, with $c$=0.0157 m/s the scale parameter and $k$=2.15 the shape parameter. In order to represent the higher variability of Titan's winds due to local gusts that are not captured with the low-resolution GCM grid (typical size of the spatial grid: 500 km$\times$170 km), we have also considered the case of an increase of wind speed by 20$\%$ (in orange in the figure). This arbitrary value corresponds to a significant increase of wind and leads to a one order of magnitude stronger sand transport. Because of the weak turbulence in the boundary layer \citep{tokano06}, it is likely to be a quite high value for high value for the representation of missing gusts. Concerning wind statistics, it is equivalent to an increase of the scale parameter of the Weibull distribution by 20$\%$ (in blue in the figure). In these conditions, the wind speed exceeds the threshold of 0.04 m/s around 0.7 $\%$ of the time (i.e. around ten times more than without correction).

\section{Simulation of methane storms}

 \textbf{Description of the TRAMS model}\\
The Titan Regional Atmospheric Modeling System (TRAMS) is a coupled, regional-scale dynamics and column microphysics model \citep{barth07,barth10}.  The governing equations for the dynamical core are the standard non-hydrostatic Reynolds-averaged primitive equations \citep{rafkin01}.  The microphysics package was adapted from Barth et al. (2006)\citep{barth06} and operates independently on each column.  Methane cloud particles form through nucleation onto submicron-sized haze particles.  Through condensation and coalescence, methane particles can grow up to millimeter sizes.   Depending on the temperature of the surrounding environment, the methane clouds form as either ice particle or droplets; melting and freezing of cloud particles is also included.  Liquid droplets are treated as a mixture of N2/CH4 following the work of Thompson et al. (1992) \citep{thompson92}.

TRAMS is run here as a 2-D model. The 2-D simulations are run with a horizontal grid spacing of 1  km, and a total horizontal extent of 1000 km.  Cyclic horizontal boundary conditions are employed.  The vertical grid extends up to about 50 km; vertical layers are more compact near the surface, starting at about 15 m spacing and extend to constant 2 km spacing above 15 km altitude. The atmosphere is initialized horizontally homogeneously using the temperature-pressure profile measured by the HASI instrument on the Huygens probe \citep{fulchignoni05}.  An initial horizontal wind is included using the $u$-velocity component from the GCM simulations.  For methane, we construct profiles using a fixed amount of convective available potential energy (CAPE); we look at cases with CAPE=250 and CAPE=500 (equivalent to a methane mixing ratio of 5 g/kg, or 10 g/kg, respectively, near the surface).  Cloud formation is initially triggered by perturbing the atmosphere with a warm bubble (the air is locally warmed in the first km), which has a maximum atmospheric temperature increase of 2 K. This 2K perturbation is a classical method to trigger deep convection in mesoscale simulations of terrestrial storm \citep{weisman88}. It is also possible to trigger deep convection with initial vertical updrafts (typically 1 m/s) but both methods lead to similar cloud dynamics \citep{hueso06}. In reality, the moist convection would be triggered by updrafts produced by planetary waves, the Hadley cell convergence, the diurnal cycle or the topography. The updraft velocities at largescale trigerring storms could be obtained from a GCM.
\\ 
\\
\textbf{Dynamics of convective cloud systems}\\
2D simulations of methane storms have been performed with our mesoscale model investigating in more details the effect of wind shear and CAPE on the storm dynamics, morphology and life time \citep{rafkin14}. It revailed behaviors similar to 2D simulations of terrestrial storms \citep{rotunno88}. We therefore expect the dynamics of the various kinds of Titan's storms (i.e. from individual convective cell to mesoscale convective systems) to be very similar to those of terrestrial storms. In particular, in our simulations the propagation and lifetime of Titan's storms is mostly controlled by the gust front. The leading edge of the gust front is able to lift moist air from the surface to trigger a new cell. When this new cell is close enough to the previous cell they merge together, increasing the lifetime of the storm and the strength of the gust front. When the gust front moves too quickly compared to the storm, they become disconnected and the storm dissipates. If a new cell were produced it would not be able to merge with the previous one. 
Fig. 1 of the letter in the manuscript reveals this behaviour. A new cell is produced in Fig. 1c at around 20 km of the pre-existing cell and merges with it. Other cells were produced by the leading edge of the gust front before this event. We found that these new cells form at 20-40 km in front of the pre-existing cell before merging with it. In figure 4d, the gust front has moved away from the mean cell. It triggers a new cell but too far (100 km) from the pre-existing cell and they dissipate.

Our idealized 2D simulations of methane storms and gust fronts are representative of large storms or mesoscale convective systems (typically 100-1000 km in latitude), as those observed by Cassini \citep{turtle11b, turtle11a}. Small isolated convective clouds, as the ones simulated in 3D by Hueso and Sanchez-Lavega 2006 \citep{hueso06}, should produce weaker gust fronts than in our simulations and with a more istropic direction. However, multiple small convective clouds should merge into larger storms (see supplementary figure 2 in Hueso and Sanchez-Lavega 2006) associated with stronger gust fronts and with a spanwise dimension similar to the one of these cloud systems (i.e. 100-1000 km). 
We also expect that some large storms or mesoscale convective systems on Titan evolve to produce squall lines or bow echos as they generally do on Earth \citep{bluestein85,weisman88}.

On Earth, a mesoscale convective system has a 3D structure, yet it becomes mostly 2D
when the wind shear is essentially 2D \citep{weisman88,parker04, coniglio06, mahoney09}. On Titan, the meridional wind shear is very weak compared to the longitudinal wind shear corresponding to the superrotation. Moreover, the Coriolis force is particularly weak at low latitude on Titan. This implies that equatorial mesocale convective systems should propagate westerly, forming fronts with no deformation by the Coriolis force. We therefore expect Titan's storm dynamics to be essentially 2D. This implies that surface winds produced by Titan's storms are primarily oriented east-west and that gust fronts should be qualitatively well represented by our idealized 2D simulations.

Finally, the direction of propagation of the gust front produced by a mesoscale convective system is essentially controlled by the transport of momentum from the mid-troposphere to the surface by cold pools \citep{mahoney09}. The presence of the superrotation necessarily implies a favoured eastward propagation for gust fronts on Titan.
A good analogy with the generation of eastward surface winds by Titan's storms is the impact of mesoscale convective systems over the Pacific warm pool during the Madden-Julian Oscillation. During such periods, mesoscale convective systems accerelate surface winds eastward in regions of fast high eastward winds \citep{mechem06, houze00, moncrieff97}.

\section{Estimation of storm frequency}
If we consider only the giant storm which was observed on 27 September 2009 \citep{turtle11a}, and if we assume that the same event occurs every equinox, we can evaluate a lower limit for the storm frequency as:
\begin{equation}
f_{min}= \frac{2 A_{\rm storm}}{A_{\rm equator} T_{\rm Titan}}
\end{equation}
$A_{\rm storm}$ is the area covered by the passage of the storm, $A_{\rm equator}$ $\approx$ 4$\times$10$^7$ km$^2$ is the area of the equatorial band (30$^\circ$ S to 30$^\circ$ N)
and $T_{\rm Titan}$ is the length of a Titan year ($\approx$29.5 terrestrial years).
The width of the storm was around 2000 km and it traveled at least 4000 km during probably a few days to produce the observed surface changes \citep{turtle11a}. Thus, thee passage of this storm covered around 20\% of Titan's equatorial band. This gives an average storm frequency of 0.4 per Titan's year or 0.2 storm per equinox. It seems reasonable to consider this value as a lower limit for the storm frequency.
Other large cloud systems have been observed during this equinox \citep{schaller09, griffith09, turtle11a, rodriguez11}. Moreover, Titan has only been observed during a short period of the equinox and not globally. Possible observations of liquid ethane on Titan's surface suggest that moderate and small rainstorms are quite common during the equinoctial season at the equator \citep{dalba12}.
We therefore expect that the mean storm frequency should rather be in the order of one storm per equinox (i.e. 2 storms per Titan year or 0.068 storms per terrestrial year).

\section{Sand flux calculations}

\textbf{Saltation threshold} 
The saltation threshold corresponds to the minimal friction speed for which the wind stress is sufficient to lift particle \citep{kok12}. It has been estimated to be around 0.04 m/s for Titan \citep{greeley85,lorenz95,kok12,lorenz13} for an optimum particle diameter of 200-300 $\mu m$ and a sediment density of around 1000 kg/m$^2$.
A simple expression of the threshold friction speed for saltation is \citep{shao00}:
\\
\begin{equation}
u_{\ast t}=A_N \sqrt{\frac{\rho_{\rm sed} - \rho_{\rm air}}{\rho_{\rm air} } g D+ \frac{\gamma}{\rho_{\rm air} D}}
\end{equation}
\\
with $A_N$=0.111 a dimensionless parameter, $\rho_{\rm air}$ the air density, $\rho_{\rm sed}$ the sediment density, $D$ the particle diameter, $g$ Titan's gravity, $\gamma$ a parameter scaling the strength of the interparticle forces. $A_N$ actually depends on the particle friction Reynolds number \citep{iversen82}, but this dependence remains limited and $A_N$ can be considered constant at first approximation \citep{shao00}.

Fig. \ref{figure_s3} shows the threshold friction speed according to particle diameter for $g$ = 1.35 m s$^{-2}$, $\rho_{\rm sed}$=1000 kg/m$^3$, $\rho_{\rm air}$=5.3 kg/m$^3$ and $\gamma$=1.5$\times$10$^{-4}$N/m. For these values, the minimal threshold friction speed is 0.045 m/s for an optimum diameter of 350 $\mu \rm m$.
For pebbles with a diameter of 1-5 cm, the threshold friction speed is 0.18-0.4 m/s. This corresponds to a wind speed at 35 m of 4-9 m/s (see below).
For dust particles with a diameter of 10 $\mu \rm m$, the threshold friction speed is 0.19 m/s. This corresponds to a wind speed at 35 m of around 4 m/s.
General circulation winds never reach such high thresholds. In contrast, gust fronts produced by methane storms may be strong enough to transport centimetric pebbles and dust particles.
Dust particles can also be raised by the impact of saltating sand, thus with a wind speed lower than threshold friction speed.
The release of dust particles could lead to the formation of dust storms that could persist in the lower atmosphere several hours or days after the passage of the methane storm. 
\begin{figure}[!h] 
\begin{center} 
	\includegraphics[width=11.5cm]{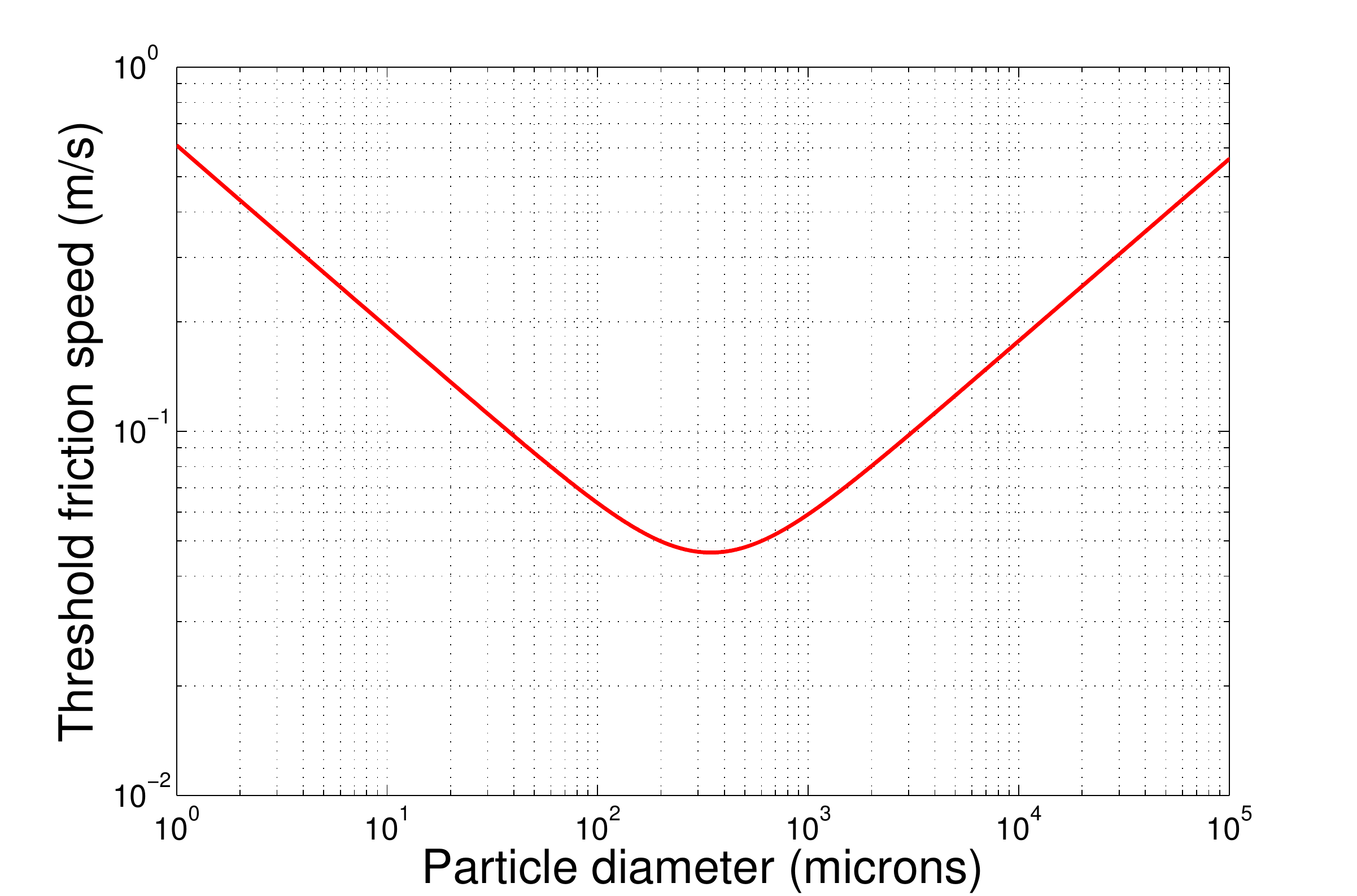}
\end{center} 
\caption{\textbf{Threshold friction speed for saltation as a function of the particle diameter.} 
The saltation speed has been calculated using relation (3) for Titan's conditions \citep{lorenz13}, with a sediment density of 1000 kg/m$^3$ and a parameter $\gamma$=1.6$\times$10$^{-4}$N/m.}
\label{figure_s3}
\end{figure} 
\\
\textbf{Sand transport law}\\
For calculating sand transport, we use the law from Kawamura (1951) \citep{kawamura51,white79,lorenz95}:
\begin{equation}
Q=2.6 \left(\frac{\rho_{\rm air}}{g \rho_{\rm sed}}\right) (u_{\ast}-u_{\ast t})(u_{\ast}+u_{\ast t})^2
\end{equation}
with $Q$ the sand flux per unit of width in m$^2$/s, $\rho_{\rm air}$ = 5.3 kg/m$^3$ the air density, $\rho_{\rm sed}$ = 1000 kg/m$^3$ the sediment density, $g$ = 1.35 m s$^{-2}$ Titan's gravity, $u_{\ast}$ the friction speed and $u_{\ast t}$ the threshold friction speed for transport.

To calculate $u_{\ast}$ from $u$, the GCM or meso-scale wind at an altitude $z$, we use the relation (1). Several other sand transport laws have been proposed \citep{kok12}. Their uses instead of the Kawamura's law could quantitively change our results (e.g. mean sand flux values) but not qualitatively (e.g. the reorientation of sand flux by methane storms).
\\
\\
\textbf{Calculation of the storm impact on the sand flux} \\
To calculate the impact of storms on the sand flux, we integrate formula (4) for the sand flux produced by the passage of one storm in our simulation and averaged it over all the domain (1000 km). In our simulations, storms are around 40 km width (in longitude) and travel around 200 km during around one day. Thus, we multiply the previous value by 5 (1000 km/200 km) to get the average impact of one storm. Finally, we divide this value by a period of half a Titan year. This yields to the mean sand flux produced by one storm per equinox (typically 0.15 m$^3$/m/year).

To obtain the total sand flux rose, we added the eastward and westward component of this sand flux, multiplied by the storm frequency per equinox, to the rose obtained with the GCM. In our mesoscale simulations, the eastward sand flux is around 2.7 times higher than the westward sand flux. The direction of surface winds produced by real storms is not likely to be purely west-east as in our 2D simulations. However, the favoured direction is eastward. Thus, the sand flux and the RDD are eastward on average.
To get more realistic figures, we added a small ad hoc angular dispersion, for the sand flux rose produced by storms: $Q\propto e^{-(\frac{\theta}{ 70^\circ})^2) }$ and $Q$ = 0 for $|\theta|$>20$^\circ$ where $\theta$ is the direction.
Fig. \ref{figure_s4} shows to the sand flux roses for different latitudes, averaging the effect of  Saturn's eccentricity. 
\\
\begin{figure}
\begin{center} 
	\includegraphics[width=5.2cm]{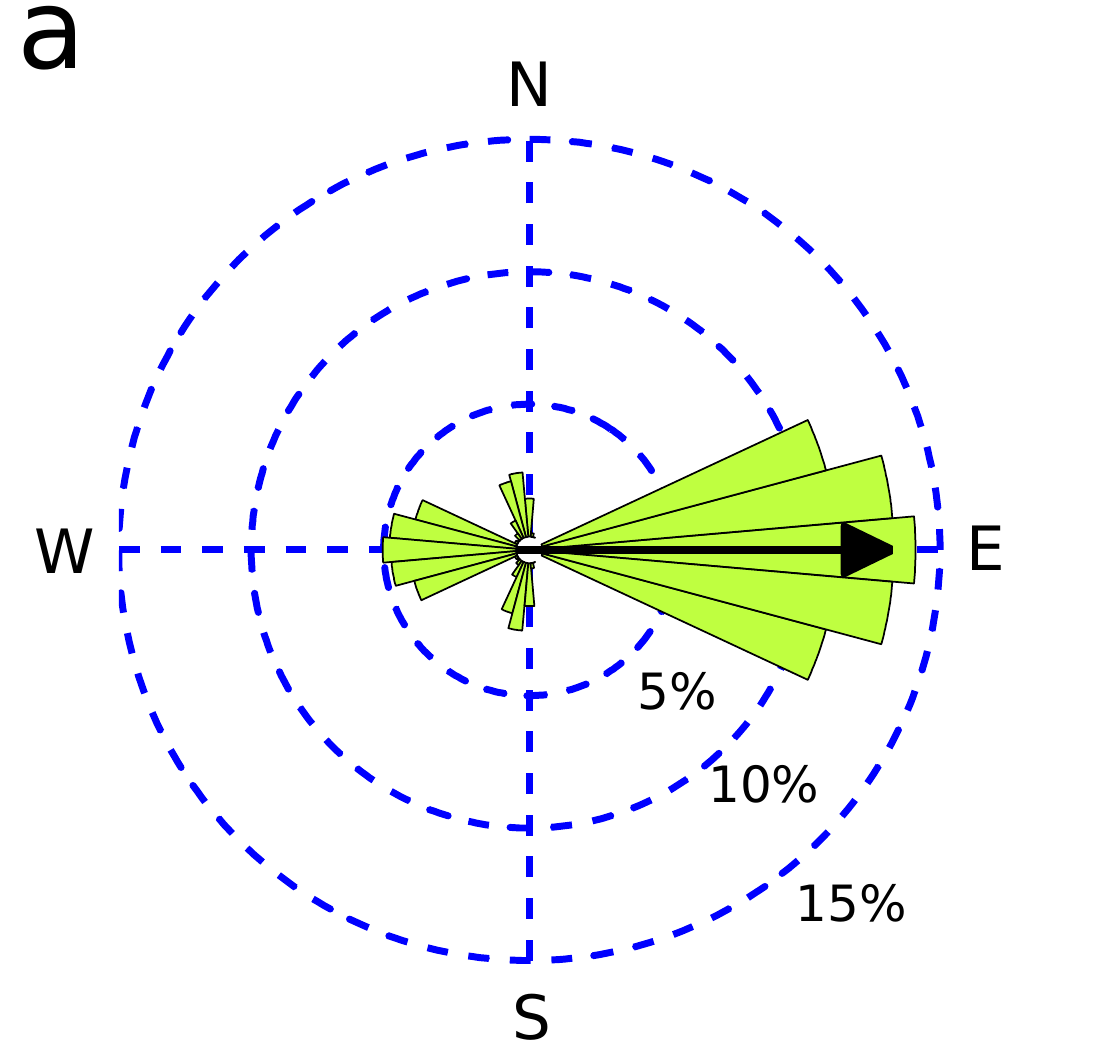}
       \includegraphics[width=5.2cm]{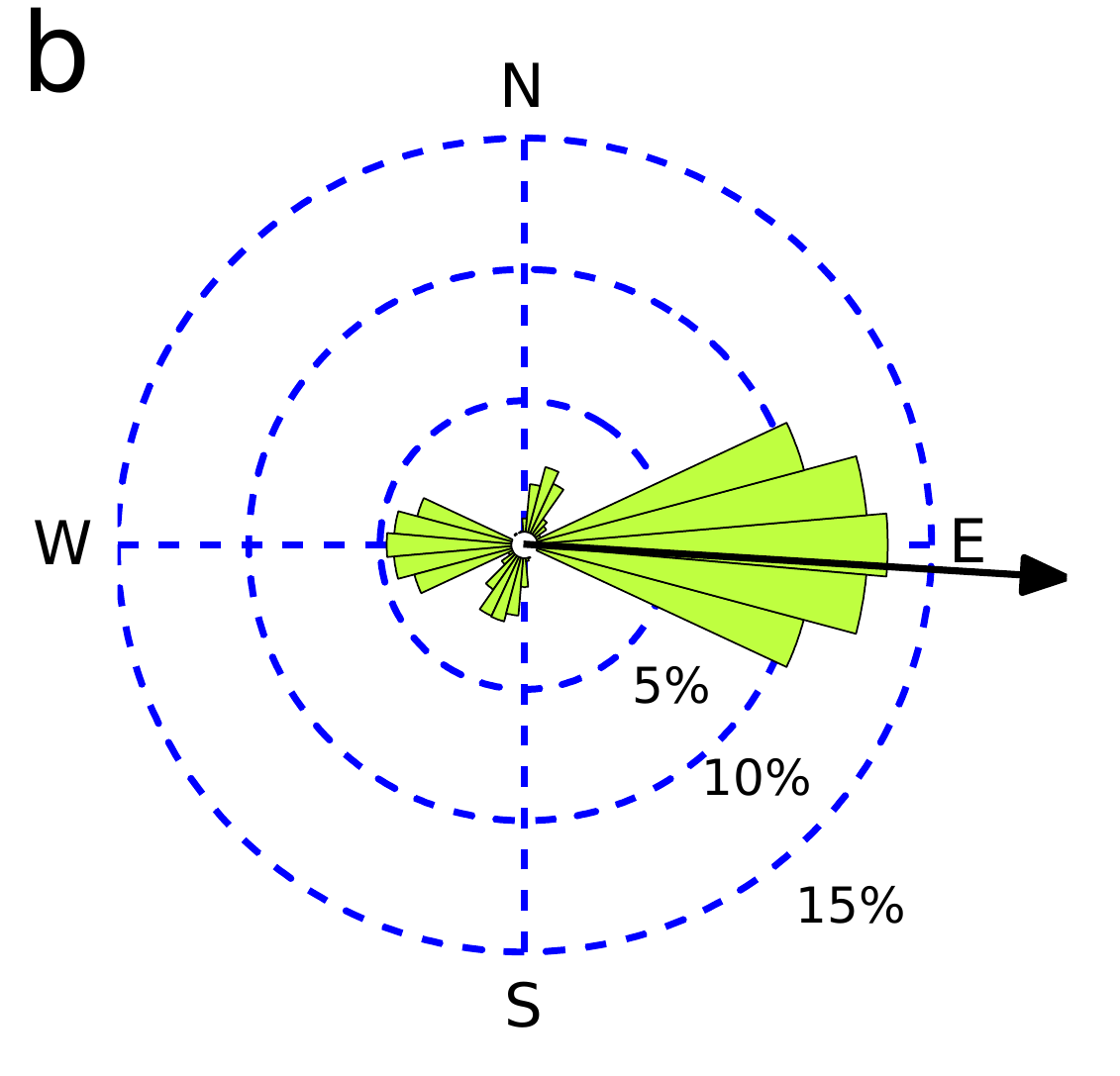}
 	\includegraphics[width=5.2cm]{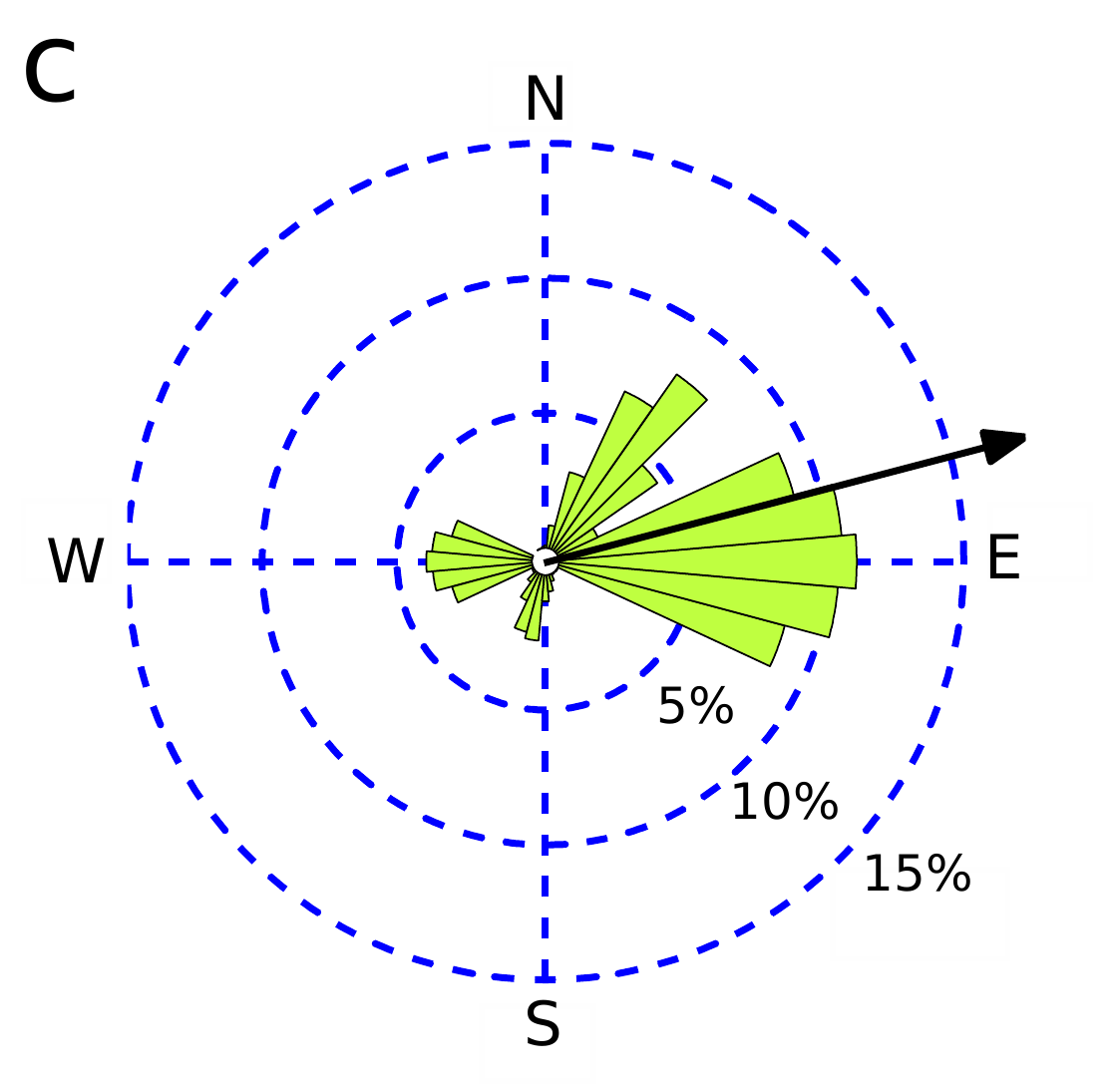}
\end{center} 
\caption{\textbf{Sand flux roses obtained by combining the GCM winds with winds produced during one typical gust front every equinox.} The arrows correspond to the resultant drift directions. a, b and c correspond to 0$^\circ$, 10$^\circ$ and 20$^\circ$ N latitudes, respectively, with GCM winds (speed increased by 20 $\%$) and a threshold of 0.04 m/s, and averaging the effect of Saturn's eccentricity.} 
\label{figure_s4}
\end{figure} 
\\
\textbf{Longitudinal dune extension rate} 
The extension rate of a longitudinal dune can be expressed as:
\begin{equation}
c=Q_{\rm RDP} \times \frac{W}{S}
\end{equation}
\\
$c$ the extension rate (in m/s), $Q_{\rm RDP}$ is the norm of the mean flux, also called the resultant drift potential (in m$^2$/s), $W$ is the width of the dune (in m) and $S$ is the cross section (in m$^2$).
For a linear dune, $S=W \times H/2$ (i.e. the surface of an isocele triangle) where $H$ is the height of the dune. We have then $c=2Q_{\rm RDP}/H$.

For Titan's dunes, $H\approx$ 100 m. If we consider the impact of one storm every equinox, $Q_{\rm RDP}$ = 0.15 m$^2$/yr and the extension rate is around 3 mm/yr.
\\
\\
\textbf{Sand flux in southern Arabia and in Egypt}
To calculate the sand flux in Rub'al-Kali desert (Fig. 4b of the letter) and in the Great Sand Sea (Fig. \ref{figure_s6}), we use the wind outputs of the ERA-Interim-project, a reanalysis from the European Centre for Medium-Range Weather Forecasts \citep{simmons06}. We analyzed the 10 m wind data for the period between the 1/1/1979 and the 31/12/2012, at 18$^\circ$N and 48$^\circ$E for Rub'al-Kali desert, and at 25.5$^\circ$N and 26.25$^\circ$E for the Great Sand Sea. Mean sand flux roses (see Fig. 4b  of the letter and Fig. \ref{figure_s6}) have been produced using these wind data and the formula (4). The mean sand flux direction (i.e. the RDD) is shown with an arrow and is parallel to the longitudinal dune orientation as predicted \citep{courrech14}.

 \begin{figure}[!h] 
\begin{center} 
 	\includegraphics[width=8.04cm]{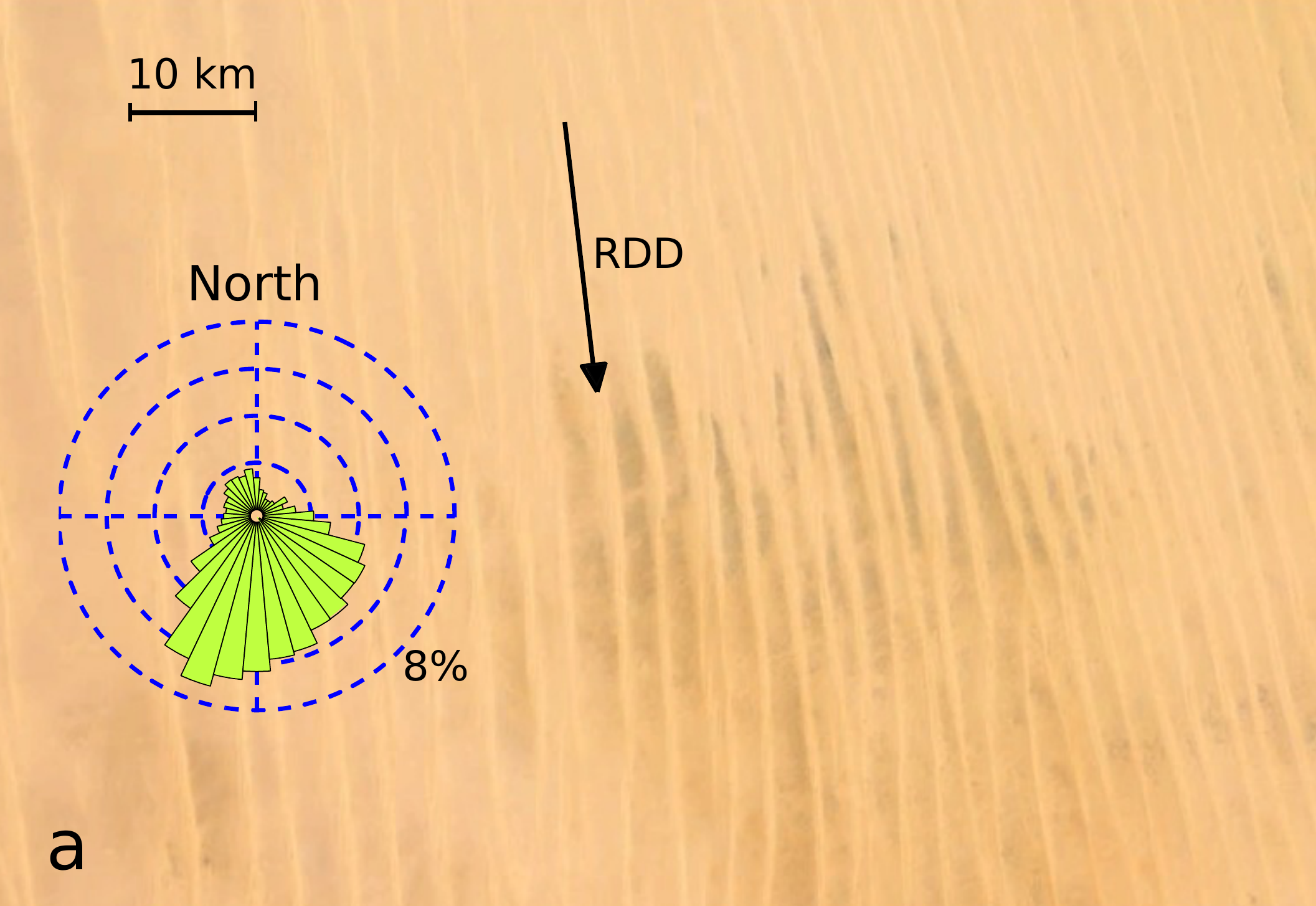}
 	\includegraphics[width=8cm]{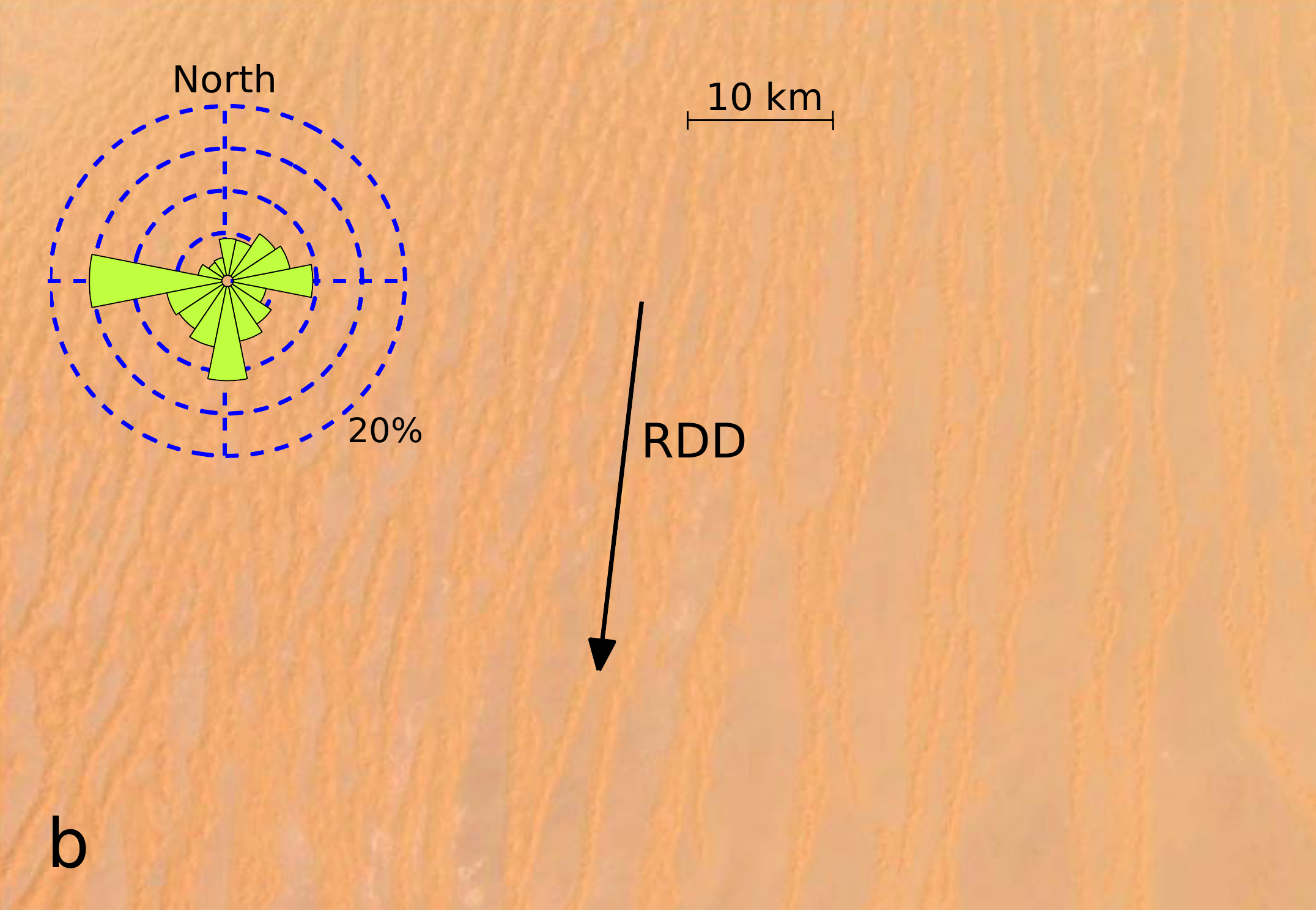}
\end{center} 
\caption{\textbf{Longitudinal dunes in Egypt and Algeria.} Map data: Google. (a) Dune field in Egypt (25.5$^\circ$N, 26.25$^\circ$E) with the sand flux roses calculated from winds at 10 m. (b) Dune field in Algeria (29.5$^\circ$ N, 5.5$^\circ$ E). The arrow corresponds to the resultant drift direction calculated with the sand flux rose.} 
\label{figure_s6}
\end{figure}


\newpage


\begin{thebibliography}{10}
\expandafter\ifx\csname url\endcsname\relax
  \def\url#1{\texttt{#1}}\fi
\expandafter\ifx\csname urlprefix\endcsname\relax\def\urlprefix{URL }\fi
\providecommand{\bibinfo}[2]{#2}
\providecommand{\eprint}[2][]{\url{#2}}

\bibitem{lorenz06}
\bibinfo{author}{{Lorenz}, R.~D.} \emph{et~al.}
\newblock \bibinfo{title}{{The Sand Seas of Titan: Cassini RADAR Observations of Longitudinal Dunes}}.
\newblock \emph{\bibinfo{journal}{Science}} \textbf{\bibinfo{volume}{312}},
  \bibinfo{pages}{724--727} (\bibinfo{year}{2006}).

\bibitem{lorenz09}
\bibinfo{author}{{Lorenz}, R.~D.} \& \bibinfo{author}{{Radebaugh}, J.}
\newblock \bibinfo{title}{{Global pattern of Titan's dunes: Radar survey from
  the Cassini prime mission}}.
\newblock \emph{\bibinfo{journal}{Geophys. Res. Lett.}}
  \textbf{\bibinfo{volume}{36}}, \bibinfo{pages}{3202} (\bibinfo{year}{2009}).

\bibitem{rodriguez14}
\bibinfo{author}{{Rodriguez}, S.} \emph{et~al.}
\newblock \bibinfo{title}{{Global mapping and characterization of Titan's dune
  fields with Cassini: correlation between RADAR and VIMS observations}}.
\newblock \emph{\bibinfo{journal}{Icarus}} \textbf{\bibinfo{volume}{230}}, \bibinfo{pages}{168-179} (\bibinfo{year}{2014}).

\bibitem{tokano08}
\bibinfo{author}{Tokano, T.}
\newblock \bibinfo{title}{{Dune-forming winds on Titan and the influence of
  topography}}.
\newblock \emph{\bibinfo{journal}{Icarus}} \textbf{\bibinfo{volume}{194}},
  \bibinfo{pages}{243--262} (\bibinfo{year}{2008}).

\bibitem{tokano10}
\bibinfo{author}{Tokano, T.}
\newblock \bibinfo{title}{{Relevance of fast westerlies at equinox for eastward
  elongation of Titan's dunes}}.
\newblock \emph{\bibinfo{journal}{Aeolian Res.}} \textbf{\bibinfo{volume}{2}},
  \bibinfo{pages}{113--127} (\bibinfo{year}{2010}).

\bibitem{barth07}
\bibinfo{author}{{Barth}, E.~L.} \& \bibinfo{author}{{Rafkin}, S.~C.~R.}
\newblock \bibinfo{title}{{TRAMS: A new dynamic cloud model for Titan's methane
  clouds}}.
\newblock \emph{\bibinfo{journal}{Geophys. Res. Lett.}}
  \textbf{\bibinfo{volume}{34}}, \bibinfo{pages}{3203} (\bibinfo{year}{2007}).

\bibitem{barth10}
\bibinfo{author}{{Barth}, E.~L.} \& \bibinfo{author}{{Rafkin}, S.~C.~R.}
\newblock \bibinfo{title}{{Convective cloud heights as a diagnostic for methane
  environment on Titan}}.
\newblock \emph{\bibinfo{journal}{Icarus}} \textbf{\bibinfo{volume}{206}},
  \bibinfo{pages}{467--484} (\bibinfo{year}{2010}).

\bibitem{lebonnois12}
\bibinfo{author}{{Lebonnois}, S.}, \bibinfo{author}{{Burgalat}, J.},
  \bibinfo{author}{{Rannou}, P.} \& \bibinfo{author}{{Charnay}, B.}
\newblock \bibinfo{title}{{Titan global climate model: A new 3-dimensional
  version of the IPSL Titan GCM}}.
\newblock \emph{\bibinfo{journal}{Icarus}} \textbf{\bibinfo{volume}{218}},
  \bibinfo{pages}{707--722} (\bibinfo{year}{2012}).

\bibitem{zhu08}
\bibinfo{author}{{Zhu}, X.}, \bibinfo{author}{{Strobel}, D.~F.} \&
  \bibinfo{author}{{Flasar}, M.~F.}
\newblock \bibinfo{title}{{Exchange of global mean angular momentum between an
  atmosphere and its underlying planet}}.
\newblock \emph{\bibinfo{journal}{Planet. \& Space Sci.}}
  \textbf{\bibinfo{volume}{56}}, \bibinfo{pages}{1524--1531}
  (\bibinfo{year}{2008}).

\bibitem{charnay12}
\bibinfo{author}{Charnay, B.} \& \bibinfo{author}{Lebonnois, S.}
\newblock \bibinfo{title}{Two boundary layers in titan's lower troposphere
  inferred from a climate model}.
\newblock \emph{\bibinfo{journal}{Nature Geoscience}}
  \textbf{\bibinfo{volume}{5}}, \bibinfo{pages}{106--109}
  (\bibinfo{year}{2012}).

\bibitem{lorenz10a}
\bibinfo{author}{Lorenz, R.~D.}, \bibinfo{author}{Claudin, P.},
  \bibinfo{author}{Andreotti, B.}, \bibinfo{author}{Radebaugh, J.} \&
  \bibinfo{author}{Tokano, T.}
\newblock \bibinfo{title}{{A 3 km atmospheric boundary layer on Titan indicated
  by dune spacing and Huygens data}}.
\newblock \emph{\bibinfo{journal}{Icarus}} \textbf{\bibinfo{volume}{205}},
  \bibinfo{pages}{719--721} (\bibinfo{year}{2010}).


\bibitem{rodriguez11}
\bibinfo{author}{Rodriguez, S.} \emph{et~al.}
\newblock \bibinfo{title}{{Titan's cloud seasonal activity from winter to spring with Cassini/VIMS}}.
\newblock \emph{\bibinfo{journal}{Icarus}} \textbf{\bibinfo{volume}{216}},
  \bibinfo{pages}{89--110} (\bibinfo{year}{2011}).

\bibitem{mitchell11}
\bibinfo{author}{{Mitchell}, J.~L.}, \bibinfo{author}{{{\'A}d{\'a}mkovics},
  M.}, \bibinfo{author}{{Caballero}, R.} \& \bibinfo{author}{{Turtle}, E.~P.}
\newblock \bibinfo{title}{{Locally enhanced precipitation organized by
  planetary-scale waves on Titan}}.
\newblock \emph{\bibinfo{journal}{Nature Geoscience}}
  \textbf{\bibinfo{volume}{4}}, \bibinfo{pages}{589--592}
  (\bibinfo{year}{2011}).

\bibitem{schneider12}
\bibinfo{author}{{Schneider}, T.}, \bibinfo{author}{{Graves}, S.~D.~B.},
  \bibinfo{author}{{Schaller}, E.~L.} \& \bibinfo{author}{{Brown}, M.~E.}
\newblock \bibinfo{title}{{Polar methane accumulation and rainstorms on Titan
  from simulations of the methane cycle}}.
\newblock \emph{\bibinfo{journal}{Nature}} \textbf{\bibinfo{volume}{481}},
  \bibinfo{pages}{58--61} (\bibinfo{year}{2012}).

\bibitem{turtle11b}
\bibinfo{author}{{Turtle}, E.~P.} \emph{et~al.}
\newblock \bibinfo{title}{{Seasonal changes in Titan's meteorology}}.
\newblock \emph{\bibinfo{journal}{Geophys. Res. Lett.}}
  \textbf{\bibinfo{volume}{38}}, \bibinfo{pages}{3203} (\bibinfo{year}{2011}).

\bibitem{turtle11a}
\bibinfo{author}{Turtle, E.~P.} \emph{et~al.}
\newblock \bibinfo{title}{{Rapid and extensive surface changes near Titan's
  equator: evidence of April showers}}.
\newblock \emph{\bibinfo{journal}{Science}} \textbf{\bibinfo{volume}{331}},
  \bibinfo{pages}{1414--1417} (\bibinfo{year}{2011}).

\bibitem{griffith09}
\bibinfo{author}{{Griffith}, C.~A.} \emph{et~al.}
\newblock \bibinfo{title}{{Characterization of Clouds in Titan's Tropical
  Atmosphere}}.
\newblock \emph{\bibinfo{journal}{Astrophys. J. l}}
  \textbf{\bibinfo{volume}{702}}, \bibinfo{pages}{L105--L109}
  (\bibinfo{year}{2009}).

\bibitem{mahoney09}
\bibinfo{author}{{Mahoney}, K.~M.}, \bibinfo{author}{{Lackmann}, G.~M.} \&
  \bibinfo{author}{{Parker}, M.~D.}
\newblock \bibinfo{title}{{The Role of Momentum Transport in the Motion of a
  Quasi-Idealized Mesoscale Convective System}}.
\newblock \emph{\bibinfo{journal}{Monthly Weather Review}}
  \textbf{\bibinfo{volume}{137}}, \bibinfo{pages}{3316} (\bibinfo{year}{2009}).

\bibitem{fulchignoni05}
\bibinfo{author}{Fulchignoni, M.} \emph{et~al.}
\newblock \bibinfo{title}{{In situ measurements of the physical characteristics
  of Titan's environment}}.
\newblock \emph{\bibinfo{journal}{Nature}} \textbf{\bibinfo{volume}{438}},
  \bibinfo{pages}{1--7} (\bibinfo{year}{2005}).

\bibitem{niemann05}
\bibinfo{author}{Niemann, H.~B.} \emph{et~al.}
\newblock \bibinfo{title}{{The abundances of constituents of Titan's atmosphere
  from the GCMS instrument on the Huygens probe}}.
\newblock \emph{\bibinfo{journal}{Nature}} \textbf{\bibinfo{volume}{438}},
  \bibinfo{pages}{1--6} (\bibinfo{year}{2005}).

\bibitem{lorenz13}
\bibinfo{author}{Lorenz, R.~D.}
\newblock \bibinfo{title}{Physics of saltation and sand transport on Titan: A
  brief review}.
\newblock \emph{\bibinfo{journal}{Icarus}}  (\bibinfo{year}{2013}).

\bibitem{burr15}
\bibinfo{author}{Burr, D.~M.} \emph{et~al.}
\newblock \bibinfo{title}{Higher-than-predicted saltation threshold wind speeds on Titan}.
\newblock \emph{\bibinfo{journal}{Nature}}  \bibinfo{volume}{517}
  \bibinfo{pages}{60--63} (\bibinfo{year}{2015}).



\bibitem{bridges12}
\bibinfo{author}{{Bridges}, N.~T.} \emph{et~al.}
\newblock \bibinfo{title}{{Earth-like sand fluxes on Mars}}.
\newblock \emph{\bibinfo{journal}{Nature}}
  \textbf{\bibinfo{volume}{485}}, \bibinfo{pages}{339-342}
  (\bibinfo{year}{2012}).



\bibitem{rubin87}
\bibinfo{author}{Rubin, D.~M.} \& \bibinfo{author}{Hunter, R.}
\newblock \bibinfo{title}{Bedform Alignment in Directionally Varying Flows}.
\newblock \emph{\bibinfo{journal}{Science}} \textbf{\bibinfo{volume}{237}},
  \bibinfo{pages}{276--278} (\bibinfo{year}{1987}).



\bibitem{courrech14}
\bibinfo{author}{Courrech~du Pont, S.}, \bibinfo{author}{Narteau, C.} \&
  \bibinfo{author}{Gao, X.}
\newblock \bibinfo{title}{Two modes for dune orientation}.
\newblock \emph{\bibinfo{journal}{Geology}} \textbf{\bibinfo{volume}{42}}, 
\bibinfo{pages}{743--746}  (\bibinfo{year}{2014}).

\bibitem{rubin09}
\bibinfo{author}{Rubin, D.~M.} \& \bibinfo{author}{Hesp, P.~A.}
\newblock \bibinfo{title}{Multiple origins of linear dunes on Earth and Titan}.
\newblock \emph{\bibinfo{journal}{Nature Geoscience}} \textbf{\bibinfo{volume}{2}},  
\bibinfo{pages}{653--658} (\bibinfo{year}{2009}).

\bibitem{radebaugh08}
\bibinfo{author}{{Radebaugh}, J.} \emph{et~al.}
\newblock \bibinfo{title}{{Dunes on Titan observed by Cassini Radar}}.
\newblock \emph{\bibinfo{journal}{Icarus}} \textbf{\bibinfo{volume}{194}},
  \bibinfo{pages}{690--703} (\bibinfo{year}{2008}).


\bibitem{ewing15}
\bibinfo{author}{Ewing, R.~C.}, \bibinfo{author}{Hayes, A.~G.}\& \bibinfo{author}{Lucas, A.}
\newblock \bibinfo{title}{Sand dune patterns on Titan controlled by long-term climate cycles}.
\newblock \emph{\bibinfo{journal}{Nature Geoscience}}  \textbf{\bibinfo{volume}{8}},
  \bibinfo{pages}{15--19} (\bibinfo{year}{2015}).

\bibitem{lucas14}
\bibinfo{author}{{Lucas}, A.} \emph{et~al.}
\newblock \bibinfo{title}{{Growth mechanisms and dune orientation on Titan}}.
\newblock \emph{\bibinfo{journal}{Geophysical Research Letters}} \textbf{\bibinfo{volume}{41}},
  \bibinfo{pages}{6093--6100} (\bibinfo{year}{2014}).


\bibitem{miller08}
\bibinfo{author}{{Miller}, S.} \emph{et~al.}
\newblock \bibinfo{title}{{Haboob dust storms of the southern Arabian Peninsula}}.
\newblock \emph{\bibinfo{journal}{J. Geophys. Res.}}
  \textbf{\bibinfo{volume}{113}}, \bibinfo{pages}{1201} (\bibinfo{year}{2008}).




\newpage
\textbf{\large Supplementary references}






\bibitem{lucas14b}
\bibinfo{author}{{Lucas}, A.} \emph{et~al.}
\newblock \bibinfo{title}{{Insights into Titan's geology and hydrology based on
  enhanced image processing of Cassini RADAR data}}.
\newblock \emph{\bibinfo{journal}{Journal of Geophysical Research}} \textbf{\bibinfo{volume}{10}},  \bibinfo{pages}{2149--2166} (\bibinfo{year}{2014}).





\bibitem{hourdin06}
\bibinfo{author}{{Hourdin}, F.} \emph{et~al.}
\newblock \bibinfo{title}{{The LMDZ4 general circulation model: climate
  performance and sensitivity to parameterized physics with emphasis on
  tropical convection}}.
\newblock \emph{\bibinfo{journal}{Clim. Dyn.}} \textbf{\bibinfo{volume}{27}},
  \bibinfo{pages}{787--813} (\bibinfo{year}{2006}).

\bibitem{tokano06b}
\bibinfo{author}{Tokano, T.}, \bibinfo{author}{Ferri, F.},
  \bibinfo{author}{Colombatti, G.}, \bibinfo{author}{{M\"akinen}, T.} \&
  \bibinfo{author}{Fulchignoni, M.}
\newblock \bibinfo{title}{{Titan's planetary boundary layer structure at the
  Huygens landing site}}.
\newblock \emph{\bibinfo{journal}{J. Geophys. Res.}}
  \textbf{\bibinfo{volume}{111}}, \bibinfo{pages}{8007} (\bibinfo{year}{2006}).

\bibitem{rannou04b}
\bibinfo{author}{Rannou, P.}, \bibinfo{author}{Hourdin, F.},
  \bibinfo{author}{McKay, C.~P.} \& \bibinfo{author}{Luz, D.}
\newblock \bibinfo{title}{{A coupled dynamics-microphysics model of Titan's
  atmosphere}}.
\newblock \emph{\bibinfo{journal}{Icarus}} \textbf{\bibinfo{volume}{170}},
  \bibinfo{pages}{443--462} (\bibinfo{year}{2004}).



\bibitem{mckay89}
\bibinfo{author}{McKay, C.~P.}, \bibinfo{author}{Pollack, J.~B.} \&
  \bibinfo{author}{Courtin, R.}
\newblock \bibinfo{title}{The thermal structure of {T}itan's atmosphere}.
\newblock \emph{\bibinfo{journal}{Icarus}} \textbf{\bibinfo{volume}{80}},
  \bibinfo{pages}{23--53} (\bibinfo{year}{1989}).

\bibitem{tokano06}
\bibinfo{author}{{Tokano}, T.} \emph{et~al.}
\newblock \bibinfo{title}{{Titan's planetary boundary layer structure at the Huygens
	landing site}}.
\newblock \emph{\bibinfo{journal}{J. Geophys. Res.}}
  \textbf{\bibinfo{volume}{111}}, \bibinfo{pages}{8007}
  (\bibinfo{year}{2006}).



\bibitem{rafkin01}
\bibinfo{author}{{Rafkin}, S.~C.~R.}, \bibinfo{author}{{Haberle}, R.~M.} \&
  \bibinfo{author}{{Michaels}, T.~I.}
\newblock \bibinfo{title}{{The Mars Regional Atmospheric Modeling System: Model
  Description and Selected Simulations}}.
\newblock \emph{\bibinfo{journal}{Icarus}} \textbf{\bibinfo{volume}{151}},
  \bibinfo{pages}{228--256} (\bibinfo{year}{2001}).

\bibitem{barth06}
\bibinfo{author}{{Barth}, E.~L.} \& \bibinfo{author}{{Toon}, O.~B.}
\newblock \bibinfo{title}{{Methane, ethane, and mixed clouds in Titan's
  atmosphere: Properties derived from microphysical modeling}}.
\newblock \emph{\bibinfo{journal}{Icarus}} \textbf{\bibinfo{volume}{182}},
  \bibinfo{pages}{230--250} (\bibinfo{year}{2006}).

\bibitem{thompson92}
\bibinfo{author}{{Thompson}, W.~R.}, \bibinfo{author}{{Zollweg}, J.~A.} \&
  \bibinfo{author}{{Gabis}, D.~H.}
\newblock \bibinfo{title}{{Vapor-liquid equilibrium thermodynamics of N2 + CH4
  - Model and Titan applications}}.
\newblock \emph{\bibinfo{journal}{Icarus}} \textbf{\bibinfo{volume}{97}},
  \bibinfo{pages}{187--199} (\bibinfo{year}{1992}).


\bibitem{weisman88}
\bibinfo{author}{{Weisman}, M.~L.}, \bibinfo{author}{{Klemp}, J.~B.} \& \bibinfo{author}{{Rotunno}, R.}
\newblock \bibinfo{title}{{Structure and Evolution of Numerically Simulated Squall Lines}}.
\newblock \emph{\bibinfo{journal}{Journal of Atmospheric Sciences}} \textbf{\bibinfo{volume}{45}},
 \bibinfo{pages}{1990--2013} (\bibinfo{year}{1988}).

\bibitem{hueso06}
\bibinfo{author}{{Hueso}, R.}, \& \bibinfo{author}{{S{\'a}nchez-Lavega}, A.}
\newblock \bibinfo{title}{{Methane storms on Saturn's moon Titan}}.
\newblock \emph{\bibinfo{journal}{Nature}} \textbf{\bibinfo{volume}{442}},
  \bibinfo{pages}{428--431} (\bibinfo{year}{2006}).

\bibitem{rafkin14}
\bibinfo{author}{{Rafkin}, S.}, \& \bibinfo{author}{{Barth}, E.}
\newblock \bibinfo{title}{{Environmental control of deep convective clouds on Titan: the combined effect of CAPE and wind shear on storm dynamics, morphology and lifetime}}.
\newblock \emph{\bibinfo{journal}{Journal of Geophysical Research, submitted}}.

\bibitem{rotunno88}
\bibinfo{author}{{Rotunno}, R.}, \bibinfo{author}{ {Klemp}, J.~B.}, \& \bibinfo{author}{{Weisman}, M.~L.}
\newblock \bibinfo{title}{{A Theory for Strong, Long-Lived Squall Lines.}}.
\newblock \emph{\bibinfo{journal}{Journal of Atmospheric Sciences}}
\textbf{\bibinfo{volume}{45}}, \bibinfo{pages}{463--485} (\bibinfo{year}{1988}).




\bibitem{bluestein85}
\bibinfo{author}{{Bluestein}, H.~B.}, \& \bibinfo{author}{{Jain}, M.~H.}
\newblock \bibinfo{title}{{Formation of Mesoscale Lines of Precipitation: Severe Squall Lines in Oklahoma during the Spring}}.
\newblock \emph{\bibinfo{journal}{Journal of Atmospheric Sciences}} \textbf{\bibinfo{volume}{42}},
  \bibinfo{pages}{1711--1732} (\bibinfo{year}{1985}).



\bibitem{parker04}
\bibinfo{author}{{Parker}, M.~D.},  \& \bibinfo{author}{{Johnson}, R.~H.}
\newblock \bibinfo{title}{{Structures and Dynamics of Quasi-2D Mesoscale Convective Systems}}.
\newblock \emph{\bibinfo{journal}{Journal of Atmospheric Sciences}} \textbf{\bibinfo{volume}{61}},
  \bibinfo{pages}{545--567} (\bibinfo{year}{2004}).


\bibitem{coniglio06}
\bibinfo{author}{{Coniglio}, M.~C.}, \bibinfo{author}{{Stensrud}, D.~J.} \& \bibinfo{author}{{Wicker}, L.~J.}
\newblock \bibinfo{title}{{Effects of Upper-Level Shear on the Structure and Maintenance of Strong Quasi-Linear Mesoscale Convective Systems}}.
\newblock \emph{\bibinfo{journal}{Journal of Atmospheric Sciences}} \textbf{\bibinfo{volume}{63}},
  \bibinfo{pages}{1231--1252} (\bibinfo{year}{2006}).




\bibitem{mechem06}
\bibinfo{author}{{Mechem}, D.~B.}, \bibinfo{author}{{Chen}, S.~S.} \& \bibinfo{author}{{Houze}, R.~A.}
\newblock \bibinfo{title}{{Momentum transport processes in the stratiform regions of mesoscale convective systems over the western Pacific warm pool}}.
\newblock \emph{\bibinfo{journal}{Quarterly Journal of the Royal Meteorological Society}} \textbf{\bibinfo{volume}{132}},
  \bibinfo{pages}{709--736} (\bibinfo{year}{2006}).


\bibitem{houze00}
\bibinfo{author}{{Houze}, Jr., R.~A.},  \bibinfo{author}{{Chen}, S.~S.},  \bibinfo{author}{{Kingsmill}, D.~E.},  \bibinfo{author}{{Serra}, Y.}, \& \bibinfo{author}{{Yuter}, S.~E.}
\newblock \bibinfo{title}{{Convection over the Pacific Warm Pool in relation to the Atmospheric Kelvin-Rossby Wave}}.
\newblock \emph{\bibinfo{journal}{Journal of Atmospheric Sciences}} \textbf{\bibinfo{volume}{57}},
  \bibinfo{pages}{3058--3089} (\bibinfo{year}{2000}).

\bibitem{moncrieff97}
\bibinfo{author}{{Moncrieff}, M.~W.}, \& \bibinfo{author}{{Klinker}, E.}
\newblock \bibinfo{title}{{Organized convective systems in the tropical western Pacific as a process in general circulation models: A TOGA COARE case-study}}.
\newblock \emph{\bibinfo{journal}{Quarterly Journal of the Royal Meteorological Society}} \textbf{\bibinfo{volume}{123}},
  \bibinfo{pages}{805--827} (\bibinfo{year}{1997}).


\bibitem{schaller09}
\bibinfo{author}{{Schaller}, E.~L.}, \bibinfo{author}{{Roe}, H.~G.},
  \bibinfo{author}{{Schneider}, T.} \& \bibinfo{author}{{Brown}, M.~E.}
\newblock \bibinfo{title}{{Storms in the tropics of Titan}}.
\newblock \emph{\bibinfo{journal}{Nature}} \textbf{\bibinfo{volume}{460}},
  \bibinfo{pages}{873--875} (\bibinfo{year}{2009}).





\bibitem{dalba12}
\bibinfo{author}{{Dalba}, P.~A.} \emph{et~al.}
\newblock \bibinfo{title}{{Cassini VIMS Observations Show Ethane is Present in
  Titan's Rainfall}}.
\newblock \emph{\bibinfo{journal}{Astrophys. J. l}}
  \textbf{\bibinfo{volume}{761}}, \bibinfo{pages}{L24} (\bibinfo{year}{2012}).






\bibitem{greeley85}
\bibinfo{author}{{Greeley}, R.} \& \bibinfo{author}{{Iversen}, J.~D.}
\newblock \bibinfo{title}{{Wind as a geological process on Earth, Mars,Venus and Titan.}}
\newblock \emph{\bibinfo{book}{Cambridge Planetary Science Series.}}
\textbf{\bibinfo{volume}{75}}
(\bibinfo{year}{1985})



\bibitem{lorenz95}
\bibinfo{author}{{Lorenz}, R.~D.}, \bibinfo{author}{{Lunine}, J.~I.},
  \bibinfo{author}{{Grier}, J.~A.} \& \bibinfo{author}{{Fisher}, M.~A.}
\newblock \bibinfo{title}{{Prediction of aeolian features on planets:
  Application to Titan paleoclimatology}}.
\newblock \emph{\bibinfo{journal}{J. Geophys. Res.}}
  \textbf{\bibinfo{volume}{100}}, \bibinfo{pages}{26377--26386}
  (\bibinfo{year}{1995}).


\bibitem{kok12}
\bibinfo{author}{{Kok}, J.~F.}, \bibinfo{author}{{Parteli}, E.~J.~R.},
  \bibinfo{author}{{Michaels}, T.~I.} \& \bibinfo{author}{{Karam}, D.~B.}
\newblock \bibinfo{title}{{The physics of wind-blown sand and dust}}.
\newblock \emph{\bibinfo{journal}{Reports on Progress in Physics}}
  \textbf{\bibinfo{volume}{75}}, \bibinfo{pages}{106901}
  (\bibinfo{year}{2012}).



\bibitem{shao00}
\bibinfo{author}{{Shao}, Y.} \& \bibinfo{author}{{Lu}, H.}
\newblock \bibinfo{title}{{A simple expression for wind erosion threshold
  friction velocity}}.
\newblock \emph{\bibinfo{journal}{Journal of Geophysical Research}}
  \textbf{\bibinfo{volume}{105}}, \bibinfo{pages}{22437}
  (\bibinfo{year}{2000}).

\bibitem{iversen82}
\bibinfo{author}{{Iversen}, J.~D.} \& \bibinfo{author}{{White}, B. R.} 
\newblock \bibinfo{title}{{Saltation threshold on Earth, Mars and Venus.}}
\newblock \emph{\bibinfo{journal}{Sedimentology}}
  \textbf{\bibinfo{volume}{29}}, \bibinfo{pages}{111-119}
 (\bibinfo{year}{1982}).

\bibitem{white79}
\bibinfo{author}{{White}, B. } 
\newblock \bibinfo{title}{{Soil transport by winds on Mars}}.
\newblock \emph{\bibinfo{journal}{Journal of Geophysical Research}}
  \textbf{\bibinfo{volume}{84}}, \bibinfo{pages}{4643-4651}
  (\bibinfo{year}{1979}).

\bibitem{kawamura51}
\bibinfo{author}{{Kawamura}, R. } 
\newblock \bibinfo{title}{{Study of sand movement by wind}}.
\newblock \emph{\bibinfo{journal}{University of California Hydraulics Engineering
Laboratory Report HEL 2-8 Berkeley}}
  (\bibinfo{year}{1951}).

\bibitem{simmons06}
\bibinfo{author}{{Simmons}, A.}, \bibinfo{author}{{Uppala}, D.} \&
  \bibinfo{author}{{Kobayashi}, S.}
\newblock \bibinfo{title}{{ERA-Interim: New ECMWF reanalysis products from 1989
  onwards}}.
\newblock \emph{\bibinfo{journal}{ECMWF newsletter}}
  \textbf{\bibinfo{volume}{110}}, \bibinfo{pages}{25--35}
  (\bibinfo{year}{2006}).




\end{thebibliography}

\end{document}